\documentclass[showpacs,floatfix,superscriptaddress]{revtex4}
\usepackage{graphicx}
\usepackage{epsfig}
\usepackage{bm}
\usepackage[utf8]{inputenc}
\usepackage{amssymb}
\usepackage{float}
\usepackage{amsmath}
\usepackage{dcolumn}
\usepackage{latexsym}
\usepackage{subfig}
\usepackage{amsthm}
\usepackage{framed}
\usepackage{cancel}
\usepackage[colorlinks]{hyperref}
\usepackage[usenames,dvipsnames]{color}
\hypersetup{
     breaklinks=true,
    pdfstartview={FitH},    
    colorlinks=true,       
    linkcolor=blue,          
    citecolor=red,        
    filecolor=magenta,      
    urlcolor=blue,           
    anchorcolor=green,      
    linktocpage=true
}
\DeclareMathOperator{\Ima}{Im}
\DeclareMathOperator{\diag}{diag}
\def\doi{http://doi.org}
\def\arxiv{http://arxiv.org/abs}

\begin{document}

\title{Hawking-Like Radiation as Tunneling from a Cosmological Black Hole in Modified Gravity: Semiclassical Approximation and Beyond}

\author{Sara Saghafi}
\email[]{s.saghafi@umz.ac.ir}
\author{Kourosh Nozari}
\email[]{knozari@umz.ac.ir (Corresponding Author)}
\affiliation{Department of Theoretical Physics, Faculty of Basic Sciences, University of Mazandaran,\\
P. O. Box 47416-95447, Babolsar, Iran}
\affiliation{ICRANet-Mazandaran, University of Mazandaran, P. O. Box 47416-95447, Babolsar, Iran}

\begin{abstract}
Hawking radiation as a quantum phenomenon is generally attributed to the existence of the event horizon of a black hole. However, we demonstrate in this paper that there is indeed ingoing Hawking-like radiation associated with apparent horizons of the first cosmological black hole solution in the framework of Scalar-Tensor-Vector Gravity (STVG) theory living in the Friedmann-Lema\^{\i}tre-Robertson-Walker (FLRW) background. Such radiation can be attributed also to the cosmological apparent horizon of the FLRW universe and even to the cosmological event horizon of de Sitter spacetime. We see how STVG theory as a successful theory for explaining black holes both on local and global scales affects the Hawking effect. Based on semiclassical approximation, we follow Hamilton-Jacobi and Parikh-Wilczek tunneling methods respectively with and without back-reaction effects. We find out that back-reaction effects make a correlation between the emission modes in Parikh-Wilczek tunneling formalism, which can address the information paradox. We obtain the corresponding Hawking-like temperature as a function of inverse powers of apparent horizons radiuses of the cosmological black hole in STVG theory. Due to non-equilibrium situation, the definition of such time-dependent temperature is effectively valid at near horizon, since Hawking (-like) radiation is a near-horizon effect. We analyze the influence of the STVG parameter associated with a deviation of the STVG theory from General Theory of Relativity (GR) on both apparent horizons and the Hawking-like temperature of the cosmological black hole. We show that increasing the STVG parameter results in appearing the Hawking-like temperature in later cosmic times with some smaller values. Also, we follow the Hamilton-Jacobi approach beyond semiclassical approximation to involve all quantum correction terms in the deduced semiclassical outcomes for the cosmological black hole in the STVG theory. Also, we prove that all results of the paper satisfy the correspondence principle so that eliminating the STVG parameter leads to achieve the corresponding results in the McVittie spacetime.
\vspace{12 pt}
\end{abstract}


\maketitle

\section{Introduction}\label{intro}

The most revolutionary theory to describe gravitational interaction is the General Theory of Relativity (GR), proposed by Albert Einstein in the early twentieth century. This theory has lots of successes in predicting and explaining astrophysical phenomena. Besides all the achievements of GR, it is not the ultimate gravitational theory. This theory needs the cosmological constant, $\Lambda$ \cite{Weinberg1989,Peebles2003} or dark energy to predict the late-time accelerated expansion of the Universe \cite{Riess1998,Garnavich1998,Perlmutter1999}. GR cannot reproduce the rotation curves of nearby galaxies \cite{Sofue2001,Sofue2016}, mass profiles of galaxy clusters \cite{Ettori2013,Voigt2006}, etc. Classically, there are two ways to modify GR -- the first one is the reformation of the mass-energy distribution of the theory on the right-hand side of Einstein equations, and the second one can be reconstruction of its geometry on the left-hand side of these equations, i.e., revising the law of gravity. Following the first way, one can consider two undetected mass-energy terms called dark matter \cite{Petraki2013} and dark energy \cite{Li2013}, which can result in the $\Lambda$ Cold Dark Matter ($\Lambda$CDM) model \cite{Weinberg2008} by taking the cosmological constant, $\Lambda$ into account as the simplest candidate for dark energy. Following the second way, one can consider several methods for changing the geometric structure of the theory, one of which is the relativistic theory of Scalar-Tensor-Vector Gravity (STVG), also known as MOdified Gravity (MOG) \cite{Moffat2006} developed by Moffat in 2006, which modifies the right-hand side of Einstein equations, automatically.

In the framework of STVG theory, gravitational effects on the fabric of spacetime are expressed by three scalar fields and a massive vector field in addition to a metric tensor field. The vector field can produce an effective repulsive gravitational force. The dynamical scalar fields in STVG theory are the mass $\tilde{\mu}$ of the vector field, the enhanced (effective) gravitational constant $G$, and the vector field coupling $\xi$. In the weak field regime, the modified acceleration law of the theory has a repulsive Yukawa force with the gravitational charge $Q=M\sqrt{\alpha G_{N}}$ added to the Newtonian acceleration law, and an enhanced Newtonian parameter $G=G_{N}(1+\alpha)$ (for more details see Ref. \cite{Moffat2006}) where $G_{N}$ is the Newtonian gravitational constant, and $\alpha$ is a free dimensionless parameter depended on the source mass-energy distribution, $M$. The STVG theory describes the rotation curves of many galaxies and the dynamics of galactic clusters without dark matter \cite{Brownstein2006a,Brownstein2006b,Brownstein2007,Moffat2013,Moffat2014,Moffat2015a}. Also, the compatibility between the STVG theory and Planck 2018 data is demonstrated in Ref. \cite{Moffat2021a}, so that reproducing the angular power spectrum features by the theory was confirmed approximately. On the other hand, the ``external field effect'' (EFE) arising from the fact that the STVG theory does not satisfy the shell or Birkhoff's theorem is studied in Ref. \cite{Moffat2021b}. Additionally, the shadow behavior as an observational signature of Kerr-MOG black hole and supermassive black holes in STVG setup, which can be examined by the Event Horizon Telescope is explored in Refs. \cite{Guo2018,Moffat2020}.

Many works in the literature focused on the characteristics and various solutions of the field equations of the STVG theory. Non-rotating and rotating black hole solutions describing the final stage of the gravitational collapse of a compact object in the STVG theory are derived in Ref. \cite{Moffat2015b}. Also, the non-rotating $D$-dimensional black hole solution of STVG field equations is studied in Ref. \cite{Cai2021}. On the other hand, the cosmological solutions of field equations of the STVG theory are investigated in Refs. \cite{Roshan2015,Jamali2018,Davari2021}. Also, the solution of the STVG field equations representing an inhomogeneity embedded in a Friedmann-Lema\^{\i}tre-Robertson-Walker (FLRW) background, which is indeed a cosmological black hole solution in STVG theory is derived in Ref. \cite{Perez2019}. In this paper, we plan to focus on the latter solution. Generally, McVittie spacetime \cite{McVittie1933} in GR was the first solution expressing an inhomogeneity embedded in a FLRW background, which is extensively investigated in Refs. \cite{Faraoni2007,Kaloper2010,Carrera2010,Faraoni2014,Nolan2014,Faraoni2015,Antoniou2016,Akbar2017} and references therein. Studying the inhomogeneous spacetimes in various theories of gravity shows us that it is required to consider the cosmological expansion of the Universe in modeling the evolution of the structures. On the other hand, both McVittie spacetime \cite{McVittie1933} and the cosmological black hole solution in STVG \cite{Perez2019} are the candidates for describing the gravitational fields of spherically symmetric mass distributions in expanding FLRW spacetime in theories of GR and STVG, respectively \cite{Kaloper2010}.

In 1973, Jacob Bekenstein \cite{Bekenstein1973} proposed that a black hole has an entropy proportional to the area of its event horizon. In the next year, Stephen Hawking showed that black holes are indeed ``radiating holes'', causing them not black \cite{Hawking1974}. Then, in 1975, in the seminal work \cite{Hawking1975} he proved that black holes are some black body objects radiating a thermal emission, known as Hawking radiation from their event horizon with a temperature, known as Hawking temperature proportional to the event horizon surface gravity. Thus, it was found that the concept of event horizon plays a crucial role in black hole physics. Also, considering the back-reaction effects naturally leads to deviating Hawking radiation from the thermal spectrum \cite{Medved2005}. On the other hand, in 1977, Gibbons and Hawking \cite{Gibbons1977} discovered a Hawking temperature associated with the cosmological event horizon with radius $l$ in the de Sitter universe to the form of $T_{_{dS}}=\hbar/(2\pi l)$, similar to such a temperature corresponding to the event horizon of a black hole. In 2000, Parikh and Wilczek \cite{Parikh2000} represent Hawking radiation as a tunneling process of particles through the event horizon of a stationary black hole based on semiclassical approximation. Until now, lots of works focused on applying the Parikh-Wilczek approach (also known as the null geodesic method) to various black hole solutions \cite{Arzano2005,Zhang2005a,Zhang2005b,Jiang2006,Wu2006,Miao2011} and studying quantum gravity effects on it \cite{Nozari2012,Hajebrahimi2020}. Also, the thermodynamics and Hawking radiation of commutative and non-commutative MOG black holes is studied respectively in Refs. \cite{Mureika2016,Saghafi2021}. On the other hand, Srinivasan et al. \cite{Srinivasan1999,Shankaranarayanan2002,Angheben2005} proposed another method to derive Hawking radiation as tunneling process with semiclassical approximation in which the classical action of tunneling particles is calculated by the Hamilton-Jacobi equation. In both Parikh-Wilczek and Hamilton-Jacobi methods due to the semiclassical approximation, quantum corrections are generally not taken into account, because only the first semiclassical term of the tunneling particle action is considered. In 2008, Banerjee and Majhi presented a generalization to the semiclassical tunneling process \cite{Banerjee2008}. They formulated the Hamilton-Jacobi method of tunneling beyond the semiclassical approximation by expanding the action of the tunneling particle in powers of the reduced Planck constant, $\hbar$ in order to apply all the higher order terms of quantum corrections to the semiclassical outcomes. There are several works focussed on studying Hawking radiation from black hole horizon as tunneling beyond the semiclassical approximation (see, e.g., \cite{Banerjee2008bb,Banerjee2008bbb,Majhi2009,Modak2009,Banerjee2009aa,Majhi2010aa,Banerjee2010bbb}).

The global concept of an event horizon in a spacetime, however, does not locally provide the possibility of locating an event horizon associated with a dynamical spacetime at a moment. This fact makes it difficult to investigate Hawking radiation in a non-stationary black hole. Due to the quasi-locally definition of apparent horizons, however, they do not refer to the global causal structure of a spacetime \cite{Faraoni2015}. Accordingly, in pioneer works \cite{Hayward2009,Di2007} following the Hamilton-Jacobi approach, the authors studied the Hawking radiation of the apparent horizon of some non-stationary black holes. Recently, the temperature and thermodynamic of a cosmological black hole and the thermal nature of a generic null surface are studied in Refs. \cite{Bhattacharya2016,Dalui2021}. In the novel work \cite{Cai2009} following the Hamilton-Jacobi and Parikh-Wilczek approaches, the authors showed that the Hawking temperature corresponding with an inward thermal spectrum, radiated from the apparent horizon of FLRW universe is to the form $T_{_{FLRW}}=\hbar/(2\pi\tilde{r}_{_{A}})$, where $\tilde{r}_{_{A}}$ is the apparent horizon radius of FLRW spacetime. This ingoing Hawking radiation is measured by a Kodama observer inside the FLRW apparent horizon. Since this interior Kodama observer is fixed inside the apparent horizon in a time-dependent frame, the radiation cannot be pure Hawking radiation and the corresponding temperature has no ``global'' definition, but ``local'' definition associated with some defined degrees of freedom in this non-equilibrium situation. So, the term ``Hawking-like'' radiation is suitable for such a spectrum detected by this Kodama observer \cite{Zhu2009,Zhu2010}. The Kodama vector \cite{Kodama1980} of the time-dependent black holes corresponding with the Kodama observer plays the same role as a Killing vector of stationary black holes \cite{Faraoni2015}. Recently, the Hawking-like radiation as tunneling from the apparent horizon of FLRW universe beyond the semiclassical approximation is investigated in Refs. \cite{Zhu2009,Jiang2009}.

As mentioned above, the STVG theory has compatibility with a lot of cosmological and galactic observations, and it accurately describes them, even in strong gravitational fields. But how about the cosmological black holes and their Hawking-like radiation? How STVG theory makes an impact on the Hawking-like radiation associated with apparent horizons of a cosmological black hole in the FLRW background? May the corresponding Hawking-like temperature as a function of the dimensionless STVG parameter, $\alpha$, be suggested as an observational tool for distinguishing the STVG theory from GR? These questions motivate us to study the Hawking-like temperature of the apparent horizons of the cosmological black hole solution in the STVG theory \cite{Perez2019} living in the FLRW background. We apply Hamilton-Jacobi and Parikh-Wilczek methods for tunneling of massive and massless particles, respectively based on semiclassical approximation, and also we examine how considering back-reaction effects in Parikh-Wilczek method can address the information paradox through a non-vanishing correlation function. Then, we use the Hamilton-Jacobi method of tunneling beyond the semiclassical approximation for a massless scalar field as a tunneling particle to apply all higher-order quantum corrections to the previous semiclassical results. It is worth noting that all quantities deduced in the paper tend to the corresponding ones of the McVittie universe in the limit $\alpha\rightarrow 0$. In the rest of the paper, we set $c=1$, where $c$ is the speed of light. Also, all figures in the paper are plotted using the scale factor of $\Lambda$CDM model, which is $a(t)=\left(\frac{\left(1-\Omega_{\Lambda,0}\right)}{\Omega_{\Lambda,0}}\sinh\left(\frac{3}{2}
H_{0}\sqrt{\Omega_{\Lambda,0}}\,t\right)^{2}\right)^{\frac{1}{3}}$ where $H_{0}=2.27\times 10^{-18}\,\mathrm{s^{-1}}\approx 70\,\,\mathrm{km\,s^{-1}\,\,Mpc^{-1}}$ and $\Omega_{\Lambda,0}=0.7$ are the late-time Hubble and the cosmological constant density parameters, respectively.

The paper is organized as follows: In Section \ref{STVG} we briefly review the metric, features, and apparent horizons of the cosmological black hole solution in STVG setup. Next, in Section \ref{HRSA} the Hamilton-Jacobi and Parikh-Wilczek methods based on the semiclassical approximation for massive and massless particles are studied, respectively. In Parikh-Wilczek approach, we consider the back-reaction effects and investigate the correlation between the emitted modes. Then, section \ref{HRBSA} includes discussing the Hamilton-Jacobi method beyond the semiclassical approximation. Finally, in Section \ref{SaC} we end with some conclusions.

\section{Cosmological Black Hole Solution in the STVG Theory}\label{STVG}

The total action of STVG theory has four terms \cite{Moffat2006}. The first term is the well-known Einstein-Hilbert action as follows
\begin{equation}\label{eq1}
S_{GR}=\frac{1}{16\pi}\int d^{4}\tilde{x}\sqrt{-\tilde{g}}\frac{1}{G}R\,,
\end{equation}
in which $\tilde{g}$ is the determinant of the metric tensor $\tilde{g}_{\mu\nu}$ of the background spacetime, $G\left(\tilde{x}\right)$ is the enhanced Newtonian parameter as a scalar field, and $R$ is the scalar curvature. Next comes the matter action $S_{M}$ for possible matter fields. The third term is the action of a massive vector field $\phi^{\mu}$ which has the mass $\tilde{\mu}$ as follows
\begin{equation}\label{eq2}
S_{\phi}=-\int d^{4}\tilde{x}\sqrt{-\tilde{g}}\left(\frac{1}{4}B^{\mu\nu}B_{\mu\nu}+V_{1}\left(\phi\right)\right)\xi\,,
\end{equation}
in which $B_{\mu\nu}=\partial_{\mu}\phi_{\nu}-\partial_{\nu}\phi_{\mu}$ and $V_{1}\left(\phi\right)=-\frac{1}{2}\tilde{\mu}\phi^{\mu}\phi_{\mu}$ denotes the potential of the vector field with coupling parameter $\xi$. Finally, the last term is the action of scalar fields as follows
\begin{equation}\label{eq3}
S_{S}=\int d^{4}\tilde{x}\sqrt{-\tilde{g}}\bigg[\frac{1}{G^{3}}\left(\frac{1}{2}\tilde{g}^{\mu\nu}\nabla_{\mu}G\nabla_{\nu}G-V_{2}
\left(G\right)\right)+\frac{1}{\tilde{\mu}^{2}G}\left(\frac{1}{2}\tilde{g}^{\mu\nu}\nabla_{\mu}\tilde{\mu}\nabla_{\nu}
\tilde{\mu}-V_{3}(\tilde{\mu})\right)+\frac{1}{G}\left(\frac{1}{2}\tilde{g}^{\mu\nu}\nabla_{\mu}\xi
\nabla_{\nu}\xi-V_{4}(\xi)\right)\bigg]\,,
\end{equation}
where $\nabla_{\mu}$ shows the covariant derivative, $G\left(\tilde{x}\right)$, $\xi\left(\tilde{x}\right)$, and $\tilde{\mu}\left(\tilde{x}\right)$ are three scalar fields in the setup, and also $V_{2}\left(G\right)$, $V_{3}\left(\xi\right)$ and $V_{4}\left(\tilde{\mu}\right)$ are their corresponding potentials, respectively. Therefore, the total action of STVG theory is written in the form of $S_{tot}=S_{GR}+S_{M}+S_{\phi}+S_{S}$.

The total stress-energy tensor in the setup is $T^{(tot)}_{\mu\nu}=T^{(M)}_{\mu\nu}+T^{(\phi)}_{\mu\nu}+T^{(S)}_{\mu\nu}$ in which $T^{(M)}_{\mu\nu}=-\frac{2}{\sqrt{-\tilde{g}}}\frac{\delta S_{M}}{\delta\tilde{g}^{\mu\nu}}$ is the stress-energy tensor of ordinary matter distribution, and
\begin{equation}\label{eq4}
T^{(\phi)}_{\mu\nu}=-\frac{2}{\sqrt{-\tilde{g}}}\frac{\delta S_{\phi}}{\delta\tilde{g}^{\mu\nu}}=-\frac{1}{4}\left(B_{\mu}^{\,\,\,\,\sigma}B_{\nu\sigma}-\frac{1}{4}
\tilde{g}_{\mu\nu}B^{\sigma\lambda}B_{\sigma\lambda}\right)\,,
\end{equation}
shows the stress-energy tensor corresponding with the vector field when $V_{1}\left(\phi\right)=0$ \cite{Moffat2015b,Perez2019}, and finally $T^{(S)}_{\mu\nu}=-\frac{2}{\sqrt{-\tilde{g}}}\frac{\delta S_{S}}{\delta\tilde{g}^{\mu\nu}}$ denotes the stress-energy tensor of the scalar fields contribution. Varying the total action $S_{tot}$ with respect to $\tilde{g}^{\mu\nu}$ results in the STVG field equations \cite{Moffat2006} to the form of
\begin{equation}\label{eq5}
G_{\mu\nu}+G\left(\nabla^{\gamma}\nabla_{\gamma}\frac{1}{G}\tilde{g}_{\mu\nu}-\nabla_{\mu}\nabla_{\nu}\frac{1}{G}\right)=8\pi G\,T^{(tot)}_{\mu\nu}\,,
\end{equation}
where $G_{\mu\nu}=R_{\mu\nu}-\frac{1}{2}\tilde{g}_{\mu\nu}R$ is the Einstein tensor. The extra term $G\left(\nabla^{\gamma}\nabla_{\gamma}\frac{1}{G}\tilde{g}_{\mu\nu}-\nabla_{\mu}\nabla_{\nu}\frac{1}{G}\right)$ in STVG field equations \eqref{eq5} arises from boundary contributions \cite{Moffat2006}.

To derive the cosmological black hole solution in STVG setup living in a FLRW background, however, the authors in Ref. \cite{Perez2019} supposed a special situation in which $\xi\left(\tilde{x}\right)=1$ and $\tilde{\mu}\left(\tilde{x}\right)=0$. Also, they considered the weak field approximation in which $Q=M\sqrt{\alpha G_{N}}$ and $G=G_{N}\left(1+\alpha\right)$ are the gravitational charge of the repulsive Yukawa force and the enhanced Newtonian parameter, respectively in which $\alpha$ is the dimensionless STVG parameter, which modifies the nature of the gravitational field \cite{Moffat2015b,Moffat2009}. Hence, the total stress-energy tensor becomes to the form $T^{(tot)}_{\mu\nu}=T^{(M)}_{\mu\nu}+T^{(\phi)}_{\mu\nu}$ where
\begin{equation}\label{eq6}
T^{(M)}_{\mu\nu}=\left(\rho+p\right)u_{\mu}u_{\nu}+p\tilde{g}_{\mu\nu}\,,
\end{equation}
is considered as the stress-energy tensor of the cosmological perfect fluid in which $\rho$, $p$, and $u^{\mu}$ are the proper energy density, the proper pressure, and the 4-velocity of the fluid, respectively. Therefore, the STVG field equations \eqref{eq5} take the simple form $G_{\mu\nu}=8\pi G\left(T^{(M)}_{\mu\nu}+T^{(\phi)}_{\mu\nu}\right)$. Finally, the authors in Ref. \cite{Perez2019} found the line element of the cosmological black hole solution in the STVG framework located in the FLRW spacetime expressed in isotropic coordinates $\tilde{x}^{\mu}=(t,x,\theta,\varphi)$ by setting $G_{N}=1$ (for more details see Ref. \cite{Perez2019}) as follows
\begin{equation}\label{eq7}
ds^{2}=-\frac{f^{2}\left(t,x\right)}{g^{2}\left(t,x\right)}dt^{2}
+a^{2}(t)g^{2}\left(t,x\right)\left(dx^{2}+x^{2}d\Omega ^{2}\right)\,,
\end{equation}
where $t$ is cosmic time, $a(t)$ is the scale factor, and we have
\begin{equation}\label{eq8}
f\left(t,x\right)=1-\frac{M^{2}(1+\alpha)}{4a^{2}(t)x^{2}}\,,
\end{equation}
\begin{equation}\label{eq9}
g\left(t,x\right)=1+\frac{M(1+\alpha)}{a(t)x}+\frac{M^{2}(1+\alpha)}{4a^{2}(t)x^{2}}\,,
\end{equation}
in which $M$ is the central source mass. The line element on the unit 2-sphere is $d\Omega^{2}=d\theta^{2}+\sin^{2}\theta\,d\varphi^{2}$. In the limit, $a(t)\rightarrow 1$, the line element \eqref{eq7} tends to be the line element of a Schwarzschild-MOG black hole, which is written in isotropic coordinates \cite{Moffat2006,Moffat2015b}, whereas in the limit $M\rightarrow 0$, Eq. \eqref{eq7} reduces to the line element of spatially flat FLRW model. As mentioned in the previous section, for $\alpha\rightarrow 0$, the McVittie spacetime in GR is recovered. Additionally, it is worth noting that by equating the gravitational charge, $Q$ in STVG setup with an electric charge, $q$ in the charged McVittie solution as $q=Q=M\sqrt{\alpha G_{N}}$, the line element of the charged McVittie spacetime \cite{Faraoni2015} becomes mathematically the same as the line element \eqref{eq7} of the cosmological black hole solution in the setup of STVG living in the FLRW background. The line element \eqref{eq7} has a scalar curvature singularity at those $x$ values that satisfy the condition
\begin{equation}\label{eq10}
a(t)x=\frac{1}{2}M\sqrt{1+\alpha}\,.
\end{equation}
This singularity can exist from the early cosmic time values. On the other hand, we focus on the spacetime events that are in the casual future of the singularity. The surface $U(t,x)=a(t)x-\frac{1}{2}M\sqrt{1+\alpha}$ at $t=0$ is in the causal past of all these events. Hence, one can interpret it as a cosmological ``Big-Bang'' singularity. Kaloper et al. provided the same explanation for the curvature singularity of McVittie spacetime in GR \cite{Kaloper2010}.

Stationary black holes, which have metric coefficients independent of time, can be characterized by the existence of event horizons. In non-stationary spacetimes, however, it is impossible to determine the location of an event horizon for a black hole since the entire spacetime manifold tends to future infinity. Instead, we can make use of the concept of the apparent horizon. Such a horizon is defined as the boundary between those light rays that are directed outwards and moving outwards, and those directed outward but moving inward. In other words, the apparent horizon is the boundary surface (usually, 3-surface) on which the null geodesic congruences change in their convergence properties. By definition, the following two conditions $\theta_{n}=0$ and $\theta_{\ell}>0$ determine the location of apparent horizons, where $\theta_{n}$ and $\theta_{\ell}$ are the expansions of the future-directed ingoing and outgoing null geodesics congruences, respectively \cite{Faraoni2015}. The areal radius of the line element \eqref{eq7} is
\begin{equation}\label{eq11}
R(t,x)\equiv R=a(t)xg(t,x)=a(t)x\left(1+\frac{M(1+\alpha)}{a(t)x}+\frac{M^{2}(1+\alpha)}{4a^{2}(t)x^{2}}\right)\,.
\end{equation}
Due to the spherical symmetry, one can rewrite the line element of the cosmological black hole solution \eqref{eq7} in the STVG theory in terms of areal radius as follows
\begin{equation}\label{eq12}
ds^{2}=h_{jk}d\tilde{x}^{j}d\tilde{x}^{k}+R^{2}d\Omega^{2}\,,
\end{equation}
where $\tilde{x}^{j}=(t,x)$, and
\begin{equation}\label{eq13}
h_{jk}=\diag\left(-\frac{f^{2}\left(t,x\right)}{g^{2}\left(t,x\right)}\,\,,\,\,a^{2}(t)g^{2}\left(t,x\right)\right)\,.
\end{equation}
Consequently, by making use of the equation $h^{jk}\partial_{j}R\,\partial_{k}R=0$ which gives the location of apparent horizons in terms of areal radius, one can attain the apparent horizons of the line element \eqref{eq12} as the roots of the following quadratic equation
\begin{equation}\label{eq14}
H^{2}R^{4}-R^{2}+4r_{0}R-4r_{1}^{2}=0\,,
\end{equation}
where
\begin{equation}\label{eq15}
r_{0}\equiv\frac{M(1+\alpha)}{2}\,,\quad r_{1}\equiv\frac{M\sqrt{\alpha(1+\alpha)}}{2}\,,
\end{equation}
and $H=\frac{\dot{a}(t)}{a(t)}$ is the Hubble parameter in which `dot' stands for time derivative. Increasing the values of the areal radius will lead to $R\rightarrow\frac{1}{H}$ which is the value of the cosmological apparent horizon in the FLRW model. On the other hand, for $H\rightarrow 0$, Eq. \eqref{eq14} reduces to a quadratic equation whose two roots are the outer and the inner event horizons in the Schwarzschild-MOG black hole \cite{Moffat2015b}. Again, as we pointed out in the previous section, for $\alpha\rightarrow 0$, Eq. \eqref{eq14} reduces to a cubic equation that gives the apparent horizons in McVittie spacetime in GR \cite{Kaloper2010}. The fact that in the appropriate limits, the roots of Eq. \eqref{eq14} as the apparent horizons of line element \eqref{eq12} become a cosmological or a black hole event horizon is a vivid sign that the line element \eqref{eq12} is a cosmological black hole in the STVG framework. To be more precise, at late cosmic time values, the positive Hubble factor shows that the line element \eqref{eq12} is a cosmological black hole in the theory of STVG \cite{Perez2019}. Also, from Eq. \eqref{eq10} the location of the cosmological singularity in terms of areal radius \eqref{eq11} is
\begin{equation}\label{eq16}
R_{cs}=M\left(1+\alpha+\sqrt{1+\alpha}\right)\,.
\end{equation}
From Eq. \eqref{eq16} one can see that the cosmological singularity appears at some larger values of the areal radius by growing $M$ (or equivalently, increasing $\alpha$).

By solving the roots of Eq. \eqref{eq14} one can obtain the apparent horizons of the cosmological black hole solution in the STVG setup. Eq. \eqref{eq14} has three physical roots: $R_{*}$, $R_{-}$, and $R_{+}$, so that $R_{*}<R_{-}<R_{+}$. The explicit forms of these apparent horizons are
\begin{equation}\label{eq17}
R_{\pm}=\frac{\sqrt{A_{1}}}{2}\pm\frac{1}{2}\sqrt{\frac{2}{H^{2}}-A_{1}-\frac{8r_{0}}{H^{2}\sqrt{A_{1}}}}\,,
\end{equation}
and
\begin{equation}\label{eq18}
R_{*}=-\frac{\sqrt{A_{1}}}{2}+\frac{1}{2}\sqrt{\frac{2}{H^{2}}-A_{1}+\frac{8r_{0}}{H^{2}\sqrt{A_{1}}}}\,,
\end{equation}
in which we have defined
\begin{equation}\label{eq19}
A_{1}\equiv\frac{\sqrt[3]{2}\left(1-48H^{2}r_{1}^{2}\right)}{3A_{2}H^{2}}+\frac{A_{2}}{3\sqrt[3]{2}H^2}+\frac{2}{3H^{2}}\,,
\end{equation}
and
\begin{equation}\label{eq20}
A_{2}\equiv\bigg(\sqrt{\left(432H^{2}r_{0}^{2}-288H^{2}r_{1}^{2}-2\right)^{2}-4\left(1-48H^{2}r_{1}^{2}\right)^{3}}
+432H^{2}r_{0}^{2}-288H^{2}r_{1}^{2}-2\bigg)^{\frac{3}{2}}\,.
\end{equation}
\begin{figure}[htb]
\centering
\includegraphics[width=0.7\textwidth]{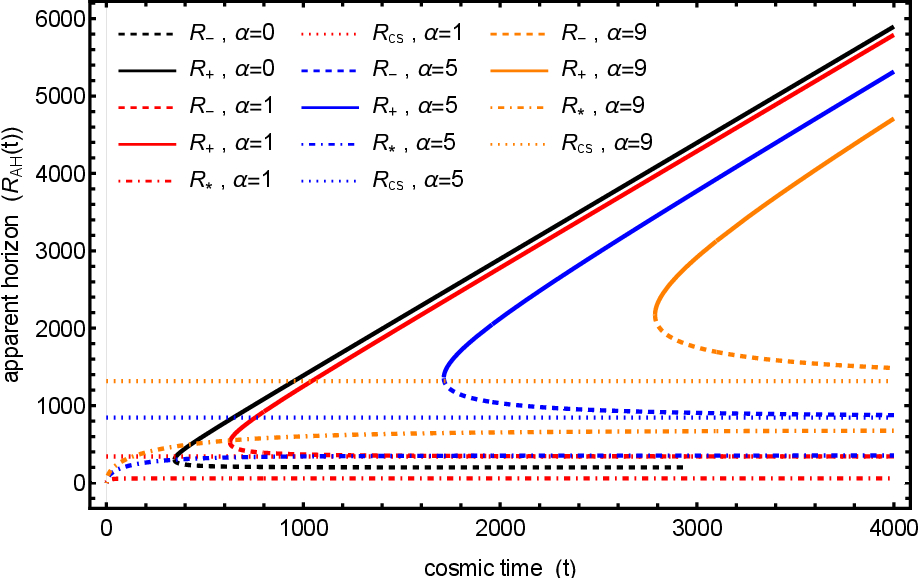}
\caption{\label{Fig1}\small{\emph{The illustration of three apparent horizons of the cosmological black hole in the STVG theory versus cosmic time $t$ for different values of $\alpha$ and two apparent horizons of McVittie solution ($\alpha=0$) in GR. The dotted lines shows the location of the cosmological singularity $R_{cs}$ for different values of $\alpha$.}}}
\end{figure}
From early values of  the cosmic time till a specific moment of it, there exist the cosmological singularity \eqref{eq16} and $R_{*}$. Thereafter, the apparent horizons $R_{-}$ and $R_{+}$ appear together in that specific value of cosmic time. Growing cosmic time $t$ results in increasing $R_{+}$, so that it reaches the value of the cosmological apparent horizon in the FLRW model. Conversely, $R_{-}$ becomes smaller by growing cosmic time, and for infinite values of cosmic time, it tends to the singularity. The apparent horizon $R_{*}$ is always inside the singularity and separated from the exterior geometry (see Fig.\ref{Fig1} and Ref. \cite{Perez2019} for more details). Accordingly, one can denote $R_{+}$ as the cosmological apparent horizon radius and $R_{-}$ as the cosmological event horizon radius of the cosmological black hole solution \eqref{eq12} in STVG theory. Fig.\ref{Fig1} is the graph of three physical apparent horizons of the cosmological black hole in STVG theory versus cosmic time $t$ for different values of $\alpha$. In this figure, the case $\alpha=0$ associated with McVittie spacetime is for comparison. From Fig.\ref{Fig1}, one can see that increasing the value of $\alpha$ leads to appear $R_{-}$ and $R_{+}$ together at some larger values of cosmic time.

\begin{figure}[htb]
\centering
\subfloat[\label{Fig2a} $R_{AH}$ versus $(t,M)$]{\includegraphics[width=0.5\textwidth]{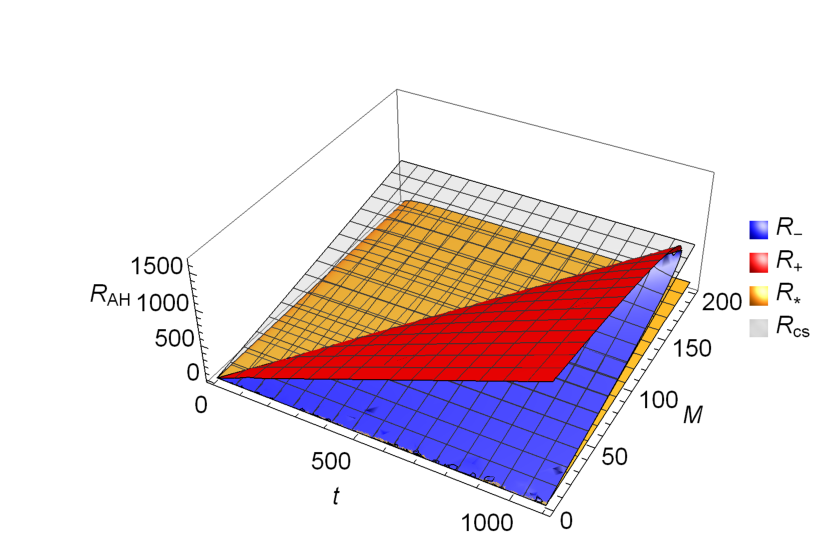}}
\,
\subfloat[\label{Fig2b} $R_{AH}$ versus $(t,\alpha)$]{\includegraphics[width=0.45\textwidth]{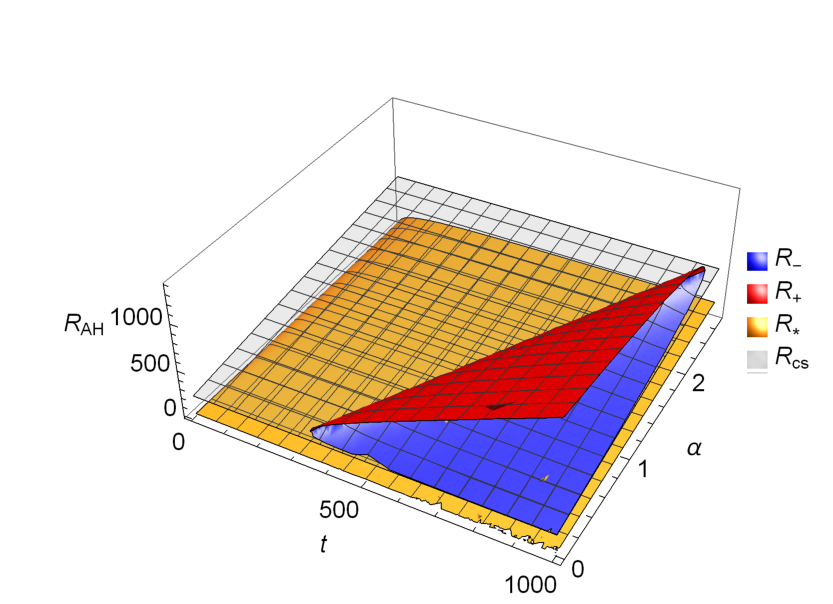}}
\caption{\label{Fig2}\small{\emph{The three-dimensional illustration of three apparent horizons $R_{-}$, $R_{+}$, and $R_{*}$ of the cosmological black hole in the STVG theory versus; (a): $(t,M)$ with $\alpha=1$ and (b): $(t,\alpha)$. Also, the gray transparent surface shows the location of the cosmological singularity, $R_{cs}$.}}}
\end{figure}
Fig.\ref{Fig2} is a three-dimensional illustration of three apparent horizons in addition to the singularity location of the cosmological black hole in the theory of STVG in terms of $(t,M)$ with $\alpha=1$ and $(t,\alpha)$. As mentioned above, Fig.\ref{Fig2} shows that the cosmological singularity will appear at some larger values of the areal radius by growing $M$ which leads to an increment in $\alpha$. From Fig.\ref{Fig2}a we see that decreasing the central mass $M$ of the black hole results in appearing $R_{-}$ and $R_{+}$ in some earlier cosmic time together, so that for $M=0$, they appear with $R_{*}$ and the cosmological singularity at $t=0$ simultaneously. Also, Fig.\ref{Fig2}b illustrates the same behavior for $R_{-}$ and $R_{+}$ in such a way that deceasing the STVG parameter $\alpha$ leads to appear $R_{-}$ and $R_{+}$ in some earlier cosmic time together, except that for $\alpha=0$ (it is associated with McVittie solution) they appear at a specific value of the cosmic time, which is not zero.

Due to the complexity of Eqs. \eqref{eq17}-\eqref{eq20} we cannot use the explicit form of these roots in the subsequent calculations to derive Hawking-like radiation as tunneling with and beyond semiclassical approximation for apparent horizons of the cosmological black hole in the setup of STVG. So, we rewrite Eq. \eqref{eq14} in the following form
\begin{equation}\label{eq21}
R^{2}\equiv R^{2}_{AH}=\frac{1}{H^{2}}\left(1-\frac{4r_{0}}{R}+\frac{4r_{1}^{2}}{R^{2}}\right)\,,
\end{equation}
in which we have made use of the following relation \cite{Perez2019}
\begin{equation}\label{eq22}
\frac{f^{2}\left(t,x\right)}{g^{2}\left(t,x\right)}=1-\frac{4r_{0}}{R}+\frac{4r_{1}^{2}}{R^{2}}\,.
\end{equation}
Eq. \eqref{eq22} is also, totaly holds in our setup. In fact, inserting Eqs. \eqref{eq11} and \eqref{eq15} into the right-hand side of Eq. \eqref{eq22}, one can simply proof this equation. Consequently, Eqs. \eqref{eq21} and \eqref{eq14} are equivalent, and  Eq. \eqref{eq21} which contains all three physical roots $R_{*}$, $R_{-}$ and $R_{+}$ can be applied to the rest of the calculations in this paper. Since we are interested in the spacetime events located in the casual future of the singularity, we focus only on $R_{-}$ and $R_{+}$ to express and plot the subsequent statements and figures.

\section{Hawking-Like Radiation as Tunneling with Semiclassical Approximation}\label{HRSA}

In this section, we confine all the calculations to the semiclassical approximation to eliminate higher-order quantum effects. In this sense, we follow both Hamilton-Jacobi \cite{Srinivasan1999,Shankaranarayanan2002,Angheben2005} and Parikh-Wilczek \cite{Parikh2000} methods in the pseudo-Painlev\'{e}-Gullstrand (PPG) coordinates to study the Hawking-like radiation as tunneling from the apparent horizons of the cosmological black hole in STVG theory.

\subsection{Hamilton-Jacobi Method for a Massive Particle}\label{HJ}

To consider the tunneling of a massive particle, we should use the coordinates system $(t, R, \theta, \varphi)$ to avoid the coordinate singularities of the spacetime metric \eqref{eq12}. Therefore, the line element \eqref{eq12} via the coordinates transformation $dx=\frac{dR}{a(t)f(t,x)}-Hxdt$ can be rewritten in the following form \cite{Perez2019}
\begin{equation}\label{eq23}
ds^{2}=-\left(1-\frac{4r_{0}}{R}+\frac{4r_{1}^{2}}{R^{2}}-H^{2}R^{2}\right)dt^{2}-\frac{2RHdR dt}{\sqrt{1-\frac{4r_{0}}{R}+\frac{4r_{1}^{2}}{R^{2}}}}
+\frac{dR^{2}}{1-\frac{4r_{0}}{R}+\frac{4r_{1}^{2}}{R^{2}}}+R^{2}d\Omega^{2}\,,
\end{equation}
which is in the PPG form \cite{Faraoni2015}. The corresponding Kodama vector \cite{Kodama1980} for the line element \eqref{eq23} is as follows
\begin{equation}\label{eq24}
K^{j}\equiv -\epsilon^{jk}\nabla_{k}R=\left(\frac{\partial}{\partial t}\right)^{j}\,,
\end{equation}
where $\epsilon_{jk}=\sqrt{|\tilde{h}|}\,(dt)_{j}\wedge(dR)_{k}$ and $\tilde{h}$ are the volume form and the determinant of the 2-metric $\tilde{h}_{jk}$ (corresponding with $(t, R)$ sector of the line element \eqref{eq23}), respectively \cite{Cai2009,Kodama1980,Faraoni2015}. Using Eqs. \eqref{eq21} and \eqref{eq22}, one can see $K_{j}K^{j}=-H^{2}R_{AH}^{2}\left(1-\frac{R^{2}}{R_{AH}^{2}}\right)$. The same result can be seen in McVittie spacetime and de Sitter solution in GR \cite{Faraoni2015}. Consequently, this Kodama vector is null at $R=R_{AH}$ and is time-like and space-like at $R<R_{AH}$ and $R>R_{AH}$, respectively. So, the deduced Kodama vector in Eq. \eqref{eq24} is time-like inside the apparent horizons in the setup, where we focus on it. Note that the existence of the Kodama vector will play a key role in this study. To see the crucial role of the Kodama vector in this setup, we discuss some differences between a stationary black hole and a time-dependent spacetime. In static and stationary situations, a time-like Killing vector field exists outside the horizon and becomes null on it. By the time-like killing vector, one can define the conserved energy of a particle moving in the stationary black hole spacetime. In dynamical situations, however, there is no time-like Killing vector, but in spherically symmetric spacetimes, the Kodama vector mimics the features of a Killing vector and gives rise to a conserved current of a particle moving in the dynamical spacetime.

To study Hawking-like radiation as tunneling via the Hamilton-Jacobi method, one can suppose a radially moving particle with mass $m$ in the spacetime background \eqref{eq23}. So, the Hamilton-Jacobi equation for the particle is as follows
\begin{equation}\label{eq25}
g^{\mu\nu}\partial_{\mu}\mathbf{S}\,\partial_{\nu}\mathbf{S}+m^{2}=0\,,
\end{equation}
in which $g_{\mu\nu}$ is the metric tensor corresponding with the line element \eqref{eq23} and $\mathbf{S}$ is the particle action. One can use the Kodama vector \eqref{eq24} to define the energy $\omega$ and radial momentum $k_{R}$ associated with this radially tunneling particle, which are measured by an observer inside the apparent horizon, called Kodama observer
\begin{equation}\label{eq26}
\omega=-K^{j}\partial_{j}\mathbf{S}=-\partial_{t}\mathbf{S}\,,\quad k_{R}=\left(\frac{\partial}{\partial R}\right)^{j}\partial_{j}\mathbf{S}=\partial_{R}\mathbf{S}\,.
\end{equation}
Therefore, the action, $\mathbf{S}$  can be written in the form of
\begin{equation}\label{eq27}
\mathbf{S}=-\int\omega dt+\int k_{R}\,dR\,.
\end{equation}
Using the action \eqref{eq27}, one can rewrite the $(t, R)$ sector of the Hamilton-Jacobi equation \eqref{eq25} to the following form
\begin{equation}\label{eq28}
\left(1-\frac{4r_{0}}{R}+\frac{4r_{1}^{2}}{R^{2}}-H^{2}R^{2}\right)k_{R}^{2}
+\frac{2HR\,\omega}{\sqrt{1-\frac{4r_{0}}{R}+\frac{4r_{1}^{2}}{R^{2}}}}k_{R} +\left(m^{2}-\frac{\omega^{2}}{1-\frac{4r_{0}}{R}+\frac{4r_{1}^{2}}{R^{2}}}\right)=0\,.
\end{equation}
Eq. \eqref{eq28} has two roots for $k_{R}$ as follow
\begin{equation}\label{eq29}
k_{R}=\frac{\frac{-HR\,\omega}{\sqrt{1-\frac{4r_{0}}{R}+\frac{4r_{1}^{2}}{R^{2}}}}\pm\sqrt{\omega^{2}-m^{2}\left(1-
\frac{4r_{0}}{R}+\frac{4r_{1}^{2}}{R^{2}}-H^{2}R^{2}\right)}}{\left(1-\frac{4r_{0}}{R}+\frac{4r_{1}^{2}}{R^{2}}\right)
\left(1-\frac{H^{2}R^{2}}{1-\frac{4r_{0}}{R}+\frac{4r_{1}^{2}}{R^{2}}}\right)}\,,
\end{equation}
in which the plus (minus) sign corresponds to the outgoing (incoming) motion. Since the energy and the radial momentum of the tunneling particle are measured by the fixed Kodama observer inside the apparent horizon, the Hawking-like radiation is, therefore, seen by the same observer, as we previously noted. The presence of the Kodama observer inside the apparent horizon necessitates us to consider the incoming motion, which is the same as the cases of the tunneling process in FLRW \cite{Cai2009} and de Sitter spacetimes \cite{Parikh2002,Medved2002}. This means that the particle tunnels from outside to inside the apparent horizon. From Eqs. \eqref{eq21} and \eqref{eq22}, we can rewrite the radial momentum \eqref{eq29} of the incoming motion as
\begin{equation}\label{eq30}
k_{R}=-\frac{\frac{R}{R_{AH}}+\sqrt{1-\frac{m^{2}}{\omega^{2}}\left(1-\frac{4r_{0}}{R}+\frac{4r_{1}^{2}}{R^{2}}\right)
\left(1-\frac{R^{2}}{R_{AH}^{2}}\right)}}{\left(1-\frac{4r_{0}}{R}+\frac{4r_{1}^{2}}{R^{2}}\right)\left(1-\frac{R^{2}}
{R_{AH}^{2}}\right)}\,\omega\,.
\end{equation}

The imaginary part of the action for the tunneling particle is
\begin{equation}\label{eq31}
\Ima\mathbf{S}=\Ima\int k_{R}\,dR\,.
\end{equation}
We take $\mathbf{S}_{in}$ as the action of incoming motion, which is the main motion of the tunneling particle in our work due to the inner Kodama observer. In fact, $\mathbf{S}_{in}$ is related to the radial momentum \eqref{eq30} of the incoming motion. So, $\Ima\mathbf{S}_{in}$ results in the ingoing emission. Whereas, $\mathbf{S}_{out}$ is associated with the radial momentum of the outgoing motion (the plus sign version of Eq. \eqref{eq29}), and $\Ima\mathbf{S}_{out}$ leads to the outgoing emission. Inserting Eq. \eqref{eq30} in Eq. \eqref{eq31} results in the following contour integral. So, we can calculate the integral through residue theorem (for more details, see Appendix \ref{app2}) to gain the imaginary part of the incoming action, $\mathbf{S}_{in}$ as follows
\begin{equation}\label{eq32}
\begin{split}
\Ima\mathbf{S}_{in} & =-\Ima\int\omega dR\frac{\frac{R}{R_{AH}}+\sqrt{1-\frac{m^{2}}{\omega^{2}}\left(1-\frac{4r_{0}}
{R}+\frac{4r_{1}^{2}}{R^{2}}\right)\left(1-\frac{R^{2}}{R_{AH}^{2}}\right)}}{\left(1-\frac{4r_{0}}{R}
+\frac{4r_{1}^{2}}{R^{2}}\right)\left(1-\frac{R^{2}}{R_{AH}^{2}}\right)}\\
& =\frac{\pi\omega R_{AH}}{1-\frac{4r_{0}}{R_{AH}}+\frac{4r_{1}^{2}}{R_{AH}^{2}}}\,.
\end{split}
\end{equation}
To prove that the tunneling process in this dynamical setup is from outside to inside the apparent horizon, we again calculate Eq. \eqref{eq31} for the outgoing action, $\mathbf{S}_{out}$ through residue theorem (for more details, see Appendix \ref{app2}) to find
\begin{equation}\label{eq33}
\begin{split}
\Ima\mathbf{S}_{out} & =-\Ima\int\omega dR\frac{\frac{R}{R_{AH}}-\sqrt{1-\frac{m^{2}}{\omega^{2}}\left(1-\frac{4r_{0}}
{R}+\frac{4r_{1}^{2}}{R^{2}}\right)\left(1-\frac{R^{2}}{R_{AH}^{2}}\right)}}{\left(1-\frac{4r_{0}}{R}
+\frac{4r_{1}^{2}}{R^{2}}\right)\left(1-\frac{R^{2}}{R_{AH}^{2}}\right)}\\
& =0\,.
\end{split}
\end{equation}
Therefore, the action of outgoing motion has no imaginary part in the dynamical setup, while in the stationary black hole spacetimes, like Schwarzschild's case in GR, the action of ingoing motion has no imaginary contribution \cite{Cai2009,Srinivasan1999,Shankaranarayanan2002,Angheben2005,Parikh2000}. Based on the semiclassical approximation, one can find the emission rate (transmission coefficient) of the tunneling particles as
\begin{equation}\label{eq34}
\Gamma\propto\exp\left(-2\Ima\mathbf{S}_{in}\right)\,.
\end{equation}
Again, by setting $\alpha=0$ in $\mathbf{S}_{in}$ one can find the corresponding emission rate for McVittie solution in GR.

Due to the similarity of Eq. \eqref{eq34} with the Boltzmann factor $\Gamma\propto\exp\left(-\frac{\omega}{T}\right),$ we see that the emission rate has the temperature $T$ in the following form
\begin{equation}\label{eq35}
T=\frac{\hbar}{2\pi R_{AH}}\left(1-\frac{4r_{0}}{R_{AH}}+\frac{4r_{1}^{2}}{R_{AH}^{2}}\right)=\frac{\hbar}{2\pi R_{AH}}\left(1-\frac{2M(1+\alpha)}{R_{AH}}+\frac{M^{2}\alpha(1+\alpha)}{R_{AH}^{2}}\right)\,.
\end{equation}
This result is independent of the mass, $m$ of the particle. Through tunneling of massive particles from outside to inside the apparent horizon, the interior Kodama observer will measure thermal radiation with temperature \eqref{eq35}. Such a procedure expressed by the Hamilton-Jacobi method can be described as thermal Hawking-like radiation of the apparent horizons of the cosmological black hole in STVG in the same meaning of the particle tunneling procedure firstly suggested by Parikh and Wilczek \cite{Parikh2000}, which provides the Hawking radiation of the black hole as a tunneling process. Again, one can see that temperature \eqref{eq35} satisfies the correspondence principle, so that by setting $\alpha=0$, we can reach the corresponding Hawking-like temperature of the apparent horizons of McVittie spacetime, which is not in the literature, and also setting $\alpha=M=0$ results in Hawking-like temperature of the apparent horizon of the spatially flat case of FLRW universe \cite{Cai2009,Zhu2010}.

So far, we found out the emission rate of the incoming action and also, the Hawking-like temperature for the ingoing Hawking-like radiation of the apparent horizons of the cosmological black hole in the STVG theory by Hamilton-Jacobi method based on semiclassical approximation. Now, we can put Eq. \eqref{eq17} of the apparent horizons $R_{-}$ and $R_{+}$ of the cosmological black hole in the STVG theory in Eqs. \eqref{eq34} and \eqref{eq35} to investigate the emission rate and Hawking-like temperature of these apparent horizons, respectively in a qualitative manner.

\begin{figure}[htb]
\centering
\includegraphics[width=0.7\textwidth]{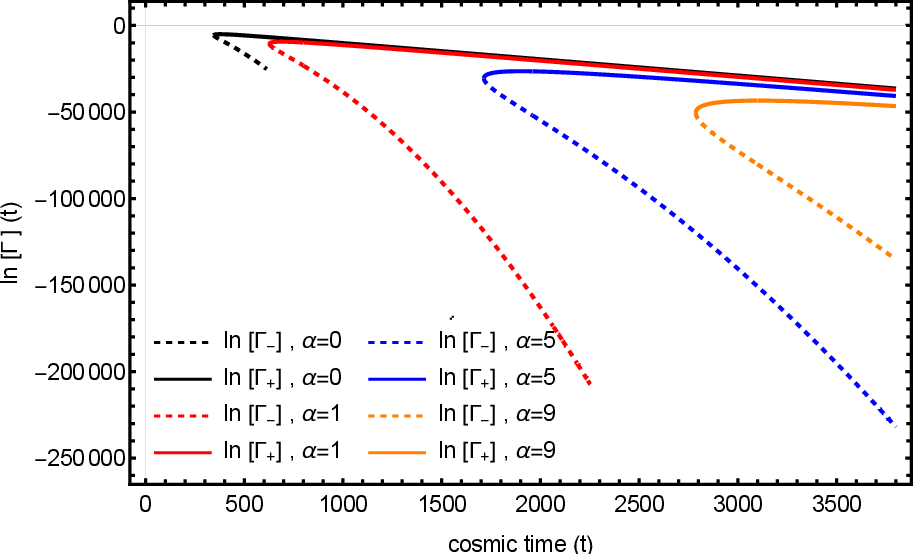}
\caption{\label{Fig3}\small{\emph{The illustration of the function $\ln\,[\Gamma]$ versus cosmic time, $t$ for the apparent horizons $R_{-}$ and $R_{+}$ of the cosmological black hole in the STVG theory for different values of $\alpha$ and McVittie solution ($\alpha=0$) in GR in which we have set $\omega=1$.}}}
\end{figure}
Fig.\ref{Fig3} illustrates the graph of the function $\ln\,[\Gamma]$ in terms of $t$ for different values of $\alpha$ associated with both apparent horizons $R_{-}$ and $R_{+}$ of the cosmological black hole in the STVG theory. Also, the case $\alpha=0$ corresponding with the McVittie solution in GR is for comparison. Fig.\ref{Fig3} shows us that for larger values of $\alpha$ the function $\ln\,[\Gamma]$ associated with $R_{-}$ and $R_{+}$ becomes available at some larger values of cosmic time in the cosmological black hole solution in STVG theory. Also, from Fig.\ref{Fig3} we see that since early cosmic time values till a specific moment of cosmic time, because the apparent horizons do not exist, the function $\ln\,[\Gamma]$ is not available. For McVittie solution in GR, the emission rate appears sooner than the cases of the cosmological black hole in STVG theory. Also, the function $\ln\,[\Gamma]$ associated with $R_{+}$ in the cosmological black hole in STVG setup tends to the corresponding case of McVittie solution in GR by increasing the cosmic time. This is because the cosmological apparent horizon $R_{+}$ in both McVittie solution in GR and the cosmological black hole in STVG framework tends to the cosmological apparent horizon in FLRW universe by increasing the cosmic time.

\begin{figure}[htb]
\centering
\includegraphics[width=0.7\textwidth]{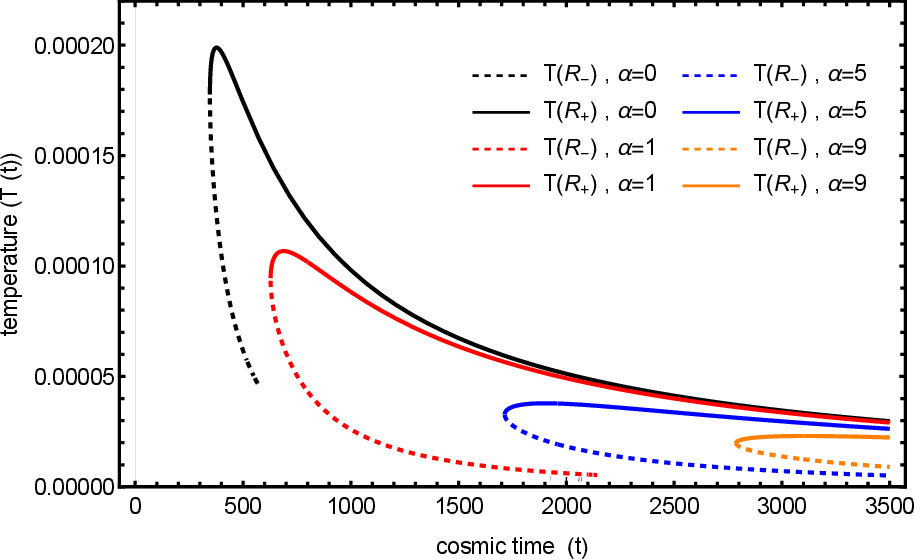}
\caption{\label{Fig4}\small{\emph{The illustration of the temperature $T$ versus cosmic time $t$ for the apparent horizons $R_{-}$ and $R_{+}$ of the cosmological black hole in the STVG theory for different values of $\alpha$ and McVittie solution in ($\alpha=0$) GR in which we have set $\hbar=1$.}}}
\end{figure}
Fig.\ref{Fig4} shows the graph of the temperature, $T$ versus cosmic time $t$ for different values of $\alpha$ associated with apparent horizons $R_{-}$ and $R_{+}$ of the cosmological black hole in the STVG theory. Also, the case $\alpha=0$ corresponding with the McVittie solution in GR is for comparison. Fig.\ref{Fig4} shows us that increasing values of $\alpha$, results in decreasing the Hawking-like temperature associated with $R_{-}$ and $R_{+}$ in the cosmological black hole solution in STVG theory, and also, they become manifest at some larger values of cosmic time. The Hawking-like temperature associated with the cosmological apparent horizon $R_{+}$ in both McVittie spacetime in GR and the cosmological black hole in STVG has some larger values, and again, the temperature in the case of the cosmological black hole in STVG, approaches the corresponding case of the McVittie solution in GR by increasing the cosmic time. Briefly, the impact of the STVG theory and its parameter, $\alpha$ on the Hawking-like temperature is to decrease the temperature. Since the parameter $\alpha$ is depend on the source mass, $M$, so in this setup, the larger the source mass, the smaller the Hawking-like temperature.

\subsection{Parikh-Wilczek Method for a Massless Particle}\label{PW}

Now, we consider back-reaction effects of a massless tunneling particle and follow the Parikh-Wilczek method \cite{Cai2009,Parikh2002,Medved2002,Parikh2000} to compute the Hawking-like temperature. The $s$-wave radiation (across the apparent horizon) of the massless tunneling particle moving along a radial null geodesics is considered to derive the Hawking-like radiation. Again, considering the semiclassical approximation, the transmission coefficient can be found as an exponential function of the imaginary part of the massless particle's action.

In the cosmological black hole spacetime described by the PPG coordinates system introduced in the line element \eqref{eq23}, the radial null geodesics can be found as follows \cite{Perez2019}
\begin{equation}\label{eq36}
\frac{dR}{dt}=\dot{R} =\pm\sqrt{1-\frac{4r_{0}}{R}+\frac{4r_{1}^{2}}{R^{2}}}\left(\sqrt{1-\frac{4r_{0}}{R}+\frac{4r_{1}^{2}}
{R^{2}}}\pm HR\right)\,.
\end{equation}
We will derive Eq. \eqref{eq36} in Appendix \ref{app1}. Again, one can rewrite Eq. \eqref{eq36} by using Eqs. \eqref{eq21} and \eqref{eq22} as
\begin{equation}\label{eq37}
\dot{R}=\pm\left(1-\frac{4r_{0}}{R}+\frac{4r_{1}^{2}}{R^{2}}\right)\left(1\pm\frac{R}{R_{AH}}\right)\,,
\end{equation}
in which the plus (minus) sign is associated with an outgoing (incoming) radial null geodesics. As previously explained, we choose the incoming mode since the particle tunneling is from outside to inside the apparent horizon.

To take into account the back-reaction effects of the massless tunneling particle in the dynamical setup, one should compute the total, physical mass-energy inside the apparent horizon. In non-stationary situations, this mass-energy inside the apparent horizon is determined with some quasi-local energy related to the apparent horizon. Due to the spherical symmetry of the cosmological black hole's line element \eqref{eq12} within the STVG theory, we can use the Misner-Sharp-Hernandez (MSH) mass \cite{Misner1964,Hernandez1966}
\begin{equation}\label{eq38}
M_{MSH}=\frac{R}{2G}\left(1-h^{jk}\partial_{j}R\,\partial_{k}R\right)\Big|_{R_{AH}}\,.
\end{equation}
One can calculate MSH mass for the cosmological black hole solution in the STVG theory by inserting Eqs. \eqref{eq12} and \eqref{eq21} into Eq. \eqref{eq38} to find the following result
\begin{equation}\label{eq39}
M_{MSH}=\frac{4}{3}\pi\rho R_{AH}^{3}+M-\frac{M^{2}\alpha}{2R_{AH}}=\frac{R_{AH}}{2(1+\alpha)}\,,
\end{equation}
in which, to derive the first term, we use the relation $H^{2}=8\pi(1+\alpha)\rho/3$ (see the proof in Ref. \cite{Perez2019}). During the instantaneous process of tunneling of a particle from the apparent horizon, we can assume that the total, physical mass-energy as the MSH mass inside the apparent horizon does not fluctuate. Thus, crossing the radiated particle with energy $\omega$ across the apparent horizon will lead to an increment of the total, physical mass-energy $M_{MSH}$ inside the apparent horizon to the amount of $M_{MSH}+\omega$. On the other hand, through tunneling, the apparent horizon's radius $R_{AH}$ will increase to $R_{AH}+\delta R_{AH}$. If we assume that $\tilde{H}$ is the Hamiltonian of the massless tunneling particle, then we can describe $d\tilde{H}$ as the energy amount crossing the apparent horizon in an infinitesimal time interval. Since we have fixed the MSH mass, we are able to write
\begin{equation}\label{eq40}
\omega=\int_{M_{MSH}}^{M_{MSH}+\omega}d\tilde{H}=\int_{R_{AH}}^{R_{AH}+\delta R_{AH}}dR_{AH}=\delta R_{AH}\,.
\end{equation}
Consequently, $R_{AH}+\omega$ can be substituted for $R_{AH}+\delta R_{AH}$ and also, $\int_{M_{MSH}}^{M_{MSH}+\omega}d\tilde{H}$ can be replaced by $\int_{0}^{\omega}d\omega'$. Therefore, one can rewrite the incoming radial null geodesics \eqref{eq37} as follows
\begin{equation}\label{eq41}
\dot{R}=-\left(1-\frac{4r_{0}}{R}+\frac{4r_{1}^{2}}{R^{2}}\right)\left(1-\frac{R}{R_{AH}+\omega}\right)\,,
\end{equation}

In the Parikh-Wilczek method, again we only need to calculate the imaginary part of the incoming action, which now can be written in the form
\begin{equation}\label{eq42}
\Ima\mathbf{S}_{in}=\Ima\int_{R_{i}}^{R_{f}}p_{_{R}}dR=\Ima\int_{R_{i}}^{R_{f}}\int_{0}^{p_{_{R}}}dp'_{_{R}}dR\,,
\end{equation}
where $p_{_{R}}$ is the canonical momentum of the tunneling particle, which is initially at $R_{i}$, somewhere slightly outside the apparent horizon, $R_{AH}$ and then crosses it to $R_{f}$, somewhere slightly inside the increased apparent horizon $R_{AH}+\delta R_{AH}$ due to the presence of back-reaction effects. The Hamiltonian equation is
\begin{equation}\label{eq43}
\dot{R}=\frac{\partial\tilde{H}}{\partial p_{_{R}}}=\frac{d\tilde{H}}{dp_{_{R}}}\bigg|_{R}\,.
\end{equation}
Finally, using Eqs. \eqref{eq40}, \eqref{eq41}, and \eqref{eq43} one can calculate the imaginary part of the incoming action \eqref{eq42} through residue theorem (for more details, see Appendix \ref{app2}) as follows
\begin{equation}\label{eq44}
\begin{split}
\Ima\mathbf{S}_{in} & =\Ima\int_{M_{MSH}}^{M_{MSH}+\omega}\int_{R_{i}}^{R_{f}}\frac{1}{\dot{R}}dR\,d\tilde{H}\\
& =-\Ima\int_{0}^{\omega}\int_{R_{i}}^{R_{f}}\frac{dR\,d\omega'}{\left(1-\frac{4r_{0}}{R}+\frac{4r_{1}^{2}}{R^{2}}\right)
\left(1-\frac{R}{R_{AH}+\omega}\right)}\\
& =\int_{0}^{\omega}\frac{\pi\left(R_{AH}+\omega\right)}{\left(1-\frac{4r_{0}}{R_{AH}+\omega}+\frac{4r_{1}^{2}}{\left(R_{AH}
+\omega\right)^{2}}\right)}d\omega'\,.
\end{split}
\end{equation}
The last integral simply can be done as
\begin{equation}\label{eq45}
\begin{split}
\Ima\mathbf{S}_{in} & =\frac{\pi}{r_{2}}\Bigg\{4r_{0}\left(4r_{0}^{2}-3r_{1}^{2}\right)\bigg(\ln\left[1+\frac{R_{AH}
-2r_{0}}{r_{2}}\right] -\ln\left[1-\frac{R_{AH}-2r_{0}}{r_{2}}\right]\bigg)\\
& +2r_{2}\Big(\left(4r_{0}^{2}-r_{1}^{2}\right)\Big(\ln\left[4r_{1}^{2}-4r_{0}(R_{AH}+\omega)
+(R_{AH}+\omega)^{2}\right]-\ln\left[4r_{1}^{2}-4r_{0}(R_{AH})+(R_{AH})^{2}\right]\Big)\Big)\\
& +2\omega\left(4r_{0}+R_{AH}+\frac{\omega}{2}\right)+4r_{0}\left(3r_{1}^{2}-4r_{0}^{2}\right)
\left(\ln\left[1+\frac{R_{AH}-2r_{0}}{r_{2}}\right]-\ln\left[1-\frac{R_{AH}-2r_{0}}{r_{2}}\right]\right)\Bigg\}\,,
\end{split}
\end{equation}
in which we have defined $r_{2}=2\sqrt{r_{0}^{2}-r_{1}^{2}}$ and the energy of the massless tunneling particle, $\omega$ is measured by the interior Kodama observer associated with the Kodama vector \eqref{eq24}.

Now, we are able to insert the imaginary part of the incoming action \eqref{eq45} into Eq. \eqref{eq34} to gain the emission rate corresponding with the ingoing Hawking-like radiation in the Parikh-Wilczek method. Then, by dropping the higher-order terms of $\omega$ and comparing the result with the Boltzmann factor based on their similarity, we can obtain the Hawking-like temperature in the Parikh-Wilczek method for the cosmological black hole in the STVG theory
\begin{equation}\label{eq46}
T\approx\frac{\hbar}{2\pi R_{AH}}\left(1-\frac{4r_{0}}{R_{AH}}+\frac{4r_{1}^{2}}{R_{AH}^{2}}\right)=\frac{\hbar}{2\pi R_{AH}}\left(1-\frac{2M(1+\alpha)}{R_{AH}}+\frac{M^{2}\alpha(1+\alpha)}{R_{AH}^{2}}\right)\,.
\end{equation}
Therefore, in semiclassical approximation, we proved that the Hawking-like temperature of the cosmological black hole solution in the STVG theory by taking into account the back-reaction effects of the massless tunneling particle up to the first order and using Parikh-Wilczek method, is the same as such a temperature for massive tunneling particle without back-reaction effects using Hamilton-Jacobi method. The temperature \eqref{eq46} completely satisfies the correspondence principle. A point to be noted is that the Hawking-like temperature of the cosmological black hole solution in the STVG theory obtained in Eq. \eqref{eq35} or \eqref{eq46} is time-dependent. So, this temperature will change over time. This is due to the non-equilibrium situation presence in this non-stationary spacetime. In Summary and Conclusions section, we will discuss about this point, in detailed.

\begin{figure}[htb]
\centering
\includegraphics[width=0.7\textwidth]{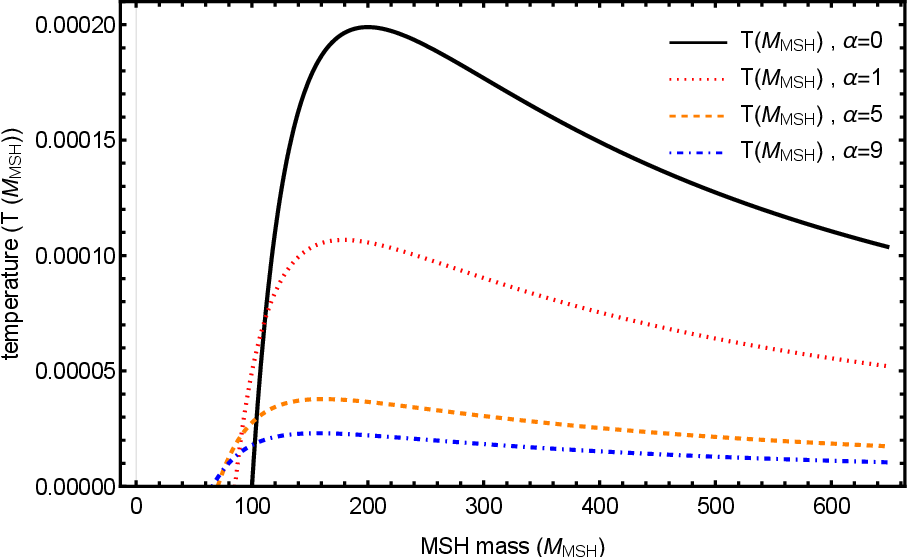}
\caption{\label{Fig5}\small{\emph{The illustration of the temperature $T$ versus the MSH mass, $M_{MSH}$, for the cosmological black hole in the STVG theory for different values of $\alpha$ and McVittie solution ($\alpha=0$) in GR in which we have set $\hbar=1$.}}}
\end{figure}
One can find the Hawking temperature \eqref{eq46} as a function of MSH mass by inserting Eq. \eqref{eq39} into Eq. \eqref{eq46} to see how increasing the quasi-local MSH mass affects the temperature. Fig.\ref{Fig5} depicts the illustration of the temperature $T$ in Eq. \eqref{eq46} versus the MSH mass, $M_{MSH}$ for different values of $\alpha$ associated with the cosmological black hole in the STVG setup. Also, the case $\alpha=0$ corresponding to the McVittie solution in GR is for comparison. Fig.\ref{Fig5} shows us that again increasing the values of $\alpha$ leads to decrease the Hawking-like temperature associated with $R_{-}$ and $R_{+}$ in the cosmological black hole solution in STVG theory. Also, from Fig.\ref{Fig5} we see that the temperature \eqref{eq46} in terms of MSH mass has no divergence in both the McVittie universe in GR and the cosmological black hole in the STVG setup.

As expected, the presence of back-reaction effects results in deviation from thermality. So, one can calculate the correlation function between radiated particles to check if the Hawking-like radiation in the Parikh-Wilczek method for the cosmological black hole in the STVG theory and in the presence of back-reaction effects is non-thermal. Generally, the correlation function is defined as $\chi(\omega_{1}+\omega_{2};\omega_{1},\omega_{2})\equiv\ln[\Gamma(\omega_{1}+\omega_{2})]-\ln[\Gamma(\omega_{1})
\Gamma(\omega_{2})]$ in which $\omega_{1}$ and $\omega_{2}$ are the energies of the radiated particles named by ``1'' and ``2'', respectively. By using Eqs. \eqref{eq34} and \eqref{eq45}, one can find the correlation function of incoming radiated particles as follows
\begin{widetext}
\begin{eqnarray}\label{eq47}
\begin{split}
\chi(\omega_{1}+\omega_{2};\,\omega_{1},\omega_{2}) & =2\pi\Bigg\{-\omega_{1}\omega_{2}+ 2\left(r_{1}^{2}-4r_{0}^{2}\right)\Big(\ln\left[-4r_{0}R_{AH}+4r_{1}^{2}+R_{AH}^{2}\right]\\
& -\ln\left[-4r_{0}(R_{AH}+\omega_{1})+4r_{1}^{2}+(R_{AH}+\omega_{1})^{2}\right]
-\ln\left[-4r_{0}(R_{AH}+\omega_{2})+4r_{1}^{2}+(R_{AH}+\omega_{2})^{2}\right]\\
& +\ln\left[-4r_{0}(R_{AH}+\omega_{1}+\omega_{2})+4r_{1}^{2}+(R_{AH}+\omega_{1}+\omega_{2})^{2}\right]\Big)
+\frac{4r_{0}\left(4r_{0}^{2}-3r_{1}^{2}\right)}{r_{2}}\\
& \times\Bigg(\ln\left[1+\frac{R_{AH}-2r_{0}}{r_{2}}\right]-\ln\left[1-\frac{R_{AH}-2r_{0}}{r_{2}}\right] +\ln\left[1-\frac{-2r_{0}+R_{AH}+\omega_{1}}{r_{2}}\right]\\
& -\ln\left[1+\frac{-2r_{0}+R_{AH}+\omega_{1}}{r_{2}}\right]+\ln\left[1-\frac{-2r_{0}+R_{AH}+
\omega_{2}}{r_{2}}\right]-\ln\left[1+\frac{-2r_{0}+R_{AH}+\omega_{2}}{r_{2}}\right]\\
& +\ln\left[1+\frac{-2r_{0}+R_{AH}+\omega_{1}+\omega_{2}}{r_{2}}\right]-
\ln\left[1-\frac{-2r_{0}+R_{AH}+\omega_{1}+\omega_{2}}{r_{2}}\right]
\Bigg)\Bigg\}\,,
\end{split}
\end{eqnarray}
\end{widetext}
which is obviously non-vanishing. So, due to the presence of back-reaction effects, the Hawking-like radiation from apparent horizons of the cosmological black hole in the STVG theory derived by Parikh-Wilczek method with semiclassical approximation deviates from thermal spectrum and the information loss problem in the setup for such an ingoing radiation has been addressed in essence.

\section{Hawking-Like Radiation as Tunneling Beyond the Semiclassical Approximation}\label{HRBSA}

In this section, we take into account all the higher-order quantum effects to study the Hawking-like radiation as tunneling beyond the semiclassical approximation for the apparent horizons of the cosmological black hole in STVG theory following the procedure introduced, e.g., in Refs. \cite{Banerjee2008,Zhu2009,Jiang2009}. To do this, we just follow the Hamilton-Jacobi method in the PPG coordinates system and expand the action of the tunneling particle in the powers of the reduced Planck constant, $\hbar$ to apply all the quantum corrections to the semiclassical approximation.

The Klein-Gordon equation for a massless scalar field $\psi$ in the background spacetime \eqref{eq23} is as follows
\begin{equation}\label{eq48}
-\frac{\hbar^{2}}{\sqrt{-g}}\,\partial_{\mu}\left(g^{\mu\nu}\sqrt{-g}\,\partial_{\nu}\right)\psi=0\,.
\end{equation}
Due to the spherical symmetry of the cosmological black hole in STVG theory, the $(t-R)$ sector of the spacetime \eqref{eq23} is considered to solve the Klein-Gordon equation \eqref{eq48}. Thus, inserting the line element \eqref{eq23} into the Klein-Gordon equation \eqref{eq48} results in the following equation
\begin{equation}\label{eq49}
\begin{split}
& -\frac{\left(\partial_{t}\right)^{2}\,\psi}{1-\frac{4r_{0}}{R}+\frac{4r_{1}^{2}}{R^{2}}}-\frac{2HR\,
\partial_{t}\,\partial_{R}\,\psi}{\left(1-\frac{4r_{0}}{R}+\frac{4r_{1}^{2}}{R^{2}}\right)^{1/2}}
+\left(-2H^{2}R+\frac{4r_{0}}{R^{2}}-\frac{8r_{1}^{2}}{R^{3}}\right)\partial_{R}\,\psi
+\left(1-\frac{4r_{0}}{R}+\frac{4r_{1}^{2}}{R^{2}}-H^{2}R^{2}\right)\left(\partial_{R}\right)^{2}\psi\\
& +\frac{H\left(\frac{4r_{0}}{R}-\frac{8r_{1}^{2}}{R^{2}}\right)\partial_{t}\,\psi}
{2\left(1-\frac{4r_{0}}{R}+\frac{4r_{1}^{2}}{R^{2}}\right)^{3/2}}-\frac{H\partial_{t}\,\psi}{\left(1-
\frac{4r_{0}}{R}+\frac{4r_{1}^{2}}{R^{2}}\right)^{1/2}}=0\,.
\end{split}
\end{equation}
The wave function of the scalar field $\psi$ is given by the WKB ansatz
\begin{equation}\label{eq50}
\psi\left(t,R\right)=\exp\left[\frac{i}{\hbar}\mathbf{S}\left(t,R\right)\right]\,,
\end{equation}
where again, $\mathbf{S}\left(t,R\right)$ is the action of the tunneling particle (i.e., the massless scalar field). Hence, adopting the ansatz \eqref{eq50} in Klein-Gordon equation \eqref{eq49} results in
\begin{equation}\label{eq51}
\begin{split}
& \frac{\partial^{2}\mathbf{S}}{\partial t^{2}}-\frac{H\left(\frac{4r_{0}}{R}-\frac{8r_{1}^{2}}{R^{2}}\right)}{2\left(1-\frac{4r_{0}}{R}
+\frac{4r_{1}^{2}}{R^{2}}\right)^{1/2}}\frac{\partial\mathbf{S}}{\partial t}-\left(1-\frac{4r_{0}}{R}
+\frac{4r_{1}^{2}}{R^{2}}\right)\left(\frac{4r_{0}}{R^{2}}-\frac{8r_{1}^{2}}{R^{3}}-2H^{2}R\right)
\frac{\partial\mathbf{S}}{\partial R}\\
& -\left(1-\frac{4r_{0}}{R}+\frac{4r_{1}^{2}}{R^{2}}\right)\left(1-\frac{4r_{0}}{R}+\frac{4r_{1}^{2}}
{R^{2}}-H^{2}R^{2}\right)\frac{\partial^{2}\mathbf{S}}{\partial R^{2}}+H\sqrt{1-\frac{4r_{0}}{R}
+\frac{4r_{1}^{2}}{R^{2}}}\frac{\partial\mathbf{S}}{\partial t}+2HR\frac{\partial^{2}\mathbf{S}}{\partial t\partial R}\\
& +\left(\frac{i}{\hbar}\right)\bigg[\left(\frac{\partial\mathbf{S}}{\partial t}\right)^{2}+2HR\frac{\partial\mathbf{S}}{\partial R}\frac{\partial\mathbf{S}}{\partial t}
-\left(1-\frac{4r_{0}}{R}+\frac{4r_{1}^{2}}{R^{2}}\right)\left(1-\frac{4r_{0}}{R}+\frac{4r_{1}^{2}}
{R^{2}}-H^{2}R^{2}\right)\left(\frac{\partial\mathbf{S}}{\partial R}\right)^{2}\bigg]=0\,.
\end{split}
\end{equation}

One can expand the action $\mathbf{S}$ of the tunneling particle in the powers of $\hbar$ as follows
\begin{equation}\label{eq52}
\begin{split}
\mathbf{S}\left(t,R\right) & =\mathbf{S}_{0}\left(t,R\right)+\hbar\,\mathbf{S}_{1}\left(t,R\right)+\hbar^{2}\mathbf{S}_{2}\left
(t,R\right)+\ldots\\
& =\mathbf{S}_{0}\left(t,R\right)+\sum_{n}\hbar^{n}\mathbf{S}_{n}\left(t,R\right)\,.
\end{split}
\end{equation}
where $n=1,2,...$ is the counter, and the semiclassical approximation is $\mathbf{S}_{0}$. So, in the semiclassical approximation, one only considers $\mathbf{S}_{0}$ while the other terms with the powers of $\hbar$ as the quantum corrections to this semiclassical value would be neglected. Hence, in this case, from \eqref{eq51} we can find
\begin{equation}\label{eq53}
\left(\frac{\partial\mathbf{S}_{0}}{\partial t}\right)^{2}+2HR\left(\frac{\partial\mathbf{S}_{0}}{\partial t}\right)\left(\frac{\partial\mathbf{S}_{0}}{\partial R}\right)-\left(1-\frac{4r_{0}}{R}+\frac{4r_{1}^{2}}{R^{2}}\right)
\left(1-\frac{4r_{0}}{R}+\frac{4r_{1}^{2}}{R^{2}}-H^{2}R^{2}\right)\left(\frac{\partial\mathbf{S}_{0}}{\partial R}\right)^{2}=0\,.
\end{equation}
Eq. \eqref{eq53} can be seen as a quadratic equation for $\frac{\partial\mathbf{S}_{0}}{\partial t}$. So, its roots can simply find as follows
\begin{equation}\label{eq54}
\frac{\partial\mathbf{S}_{0}}{\partial t}=\left[-HR\pm\sqrt{\left(1-\frac{4r_{0}}{R}+\frac{4r_{1}^{2}}{R^{2}}\right)^{2}+H^{2}R^{2}
\left(\frac{4r_{0}}{R}-\frac{4r_{1}^{2}}{R^{2}}\right)}\right]\frac{\partial\mathbf{S}_{0}}{\partial R}\,.
\end{equation}
Now, inserting Eq. \eqref{eq52} into Eq. \eqref{eq51} and using the solution \eqref{eq54}, after some lengthy calculations, we can achieve the other equations for $\frac{\partial\mathbf{S}_{n}}{\partial t}$ which their roots are as follows
\begin{equation}\label{eq55}
\begin{split}
& \hbar^{1}:\quad\frac{\partial\mathbf{S}_{1}}{\partial t}=\left[-HR\pm\sqrt{\left(1-\frac{4r_{0}}{R}+\frac{4r_{1}^{2}}{R^{2}}\right)^{2}+H^{2}R^{2}
\left(\frac{4r_{0}}{R}-\frac{4r_{1}^{2}}{R^{2}}\right)}\right]\frac{\partial\mathbf{S}_{1}}{\partial R}\,,\\
& \hbar^{2}:\quad\frac{\partial\mathbf{S}_{2}}{\partial t}=\left[-HR\pm\sqrt{\left(1-\frac{4r_{0}}{R}+\frac{4r_{1}^{2}}{R^{2}}\right)^{2}+H^{2}R^{2}
\left(\frac{4r_{0}}{R}-\frac{4r_{1}^{2}}{R^{2}}\right)}\right]\frac{\partial\mathbf{S}_{2}}{\partial R}\,,\\
& \quad\,\vdots\\
& \hbar^{n}:\quad\frac{\partial\mathbf{S}_{n}}{\partial t}=\left[-HR\pm\sqrt{\left(1-\frac{4r_{0}}{R}+\frac{4r_{1}^{2}}{R^{2}}\right)^{2}+H^{2}R^{2}
\left(\frac{4r_{0}}{R}-\frac{4r_{1}^{2}}{R^{2}}\right)}\right]\frac{\partial\mathbf{S}_{n}}{\partial R}\,.
\end{split}
\end{equation}
Since all the relations in the linear differential equations \eqref{eq54} and \eqref{eq55} for each order of $\hbar$ have the same functional form, their solutions are not independent. Hence, all $\mathbf{S}_{n}$ are proportional to $\mathbf{S}_{0}$. Therefore, one can rewrite Eq. \eqref{eq52} in the following form
\begin{equation}\label{eq56}
\mathbf{S}\left(t,R\right)=\mathbf{S}_{0}\left(t,R\right)\left(1+\sum_{n}\lambda_{n}\hbar^{n}\right)\,,
\end{equation}
where $\lambda_{n}$ are the proportionality constants. As we mentioned above, $\mathbf{S}_{0}$ is the semiclassical term and the other terms $\mathbf{S}_{0}\left(t,R\right)\left(\sum_{n}\lambda_{n}\hbar^{n}\right)$ are the higher-order quantum corrections due to the presence of $\hbar$. Based on the ansatz \eqref{eq50} and the expansion \eqref{eq52}, one can observe that $\mathbf{S}_{0}$ has the dimension of $\hbar$. Consequently, the proportionality constants $\lambda_{n}$ have the dimension of $(1/\hbar^{n})$. Since we have set $G_{N}=c=1$, the reduced Planck constant $\hbar$ has the order of the square of the Planck mass, $M_{pl}$. Thus, the constants $\lambda_{n}$ have the dimension of $\left(1/M_{MSH}^{2}\right)^{n}$. So, using \eqref{eq39} we can rewrite the action \eqref{eq56} to the form
\begin{equation}\label{eq57}
\mathbf{S}\left(t,R\right)=\mathbf{S}_{0}\left(t,R\right)\left(1+\sum_{n}\frac{4(1+\alpha)^{2}}{R_{AH}^{2}}\eta_{_{n}}\hbar^{n}
\right)\,,
\end{equation}
where $\eta_{_{n}}$ are some dimensionless proportionality constants.

As in the previous section, again the Kodama vector \eqref{eq24} is used to define the energy $\omega$ and the radial momentum $k_{R}$ of the tunneling particle in Eq. \eqref{eq26}. Therefore, one can read $\mathbf{S}_{0}$ as
\begin{equation}\label{eq58}
\mathbf{S}_{0}=-\int\omega dt+\int k_{R}dR\,.
\end{equation}
Combining Eq. \eqref{eq58} and the first relation in Eq. \eqref{eq55} results in Eq. \eqref{eq29} for the radial momentum $k_{R}$. Here we proceed with both outgoing and ingoing motion. So, using Eqs. \eqref{eq29}, \eqref{eq57}, and \eqref{eq58}, therefore, we can read, respectively, the action of the tunneling particle for both outgoing and ingoing motion to the following forms
\begin{equation}\label{eq59}
\mathbf{S}_{out}\left(t,R\right)=\left[1+\sum_{n}\frac{4(1+\alpha)^{2}}{R_{AH}^{2}}\eta_{_{n}}\hbar^{n}\right]\left[-\int\omega dt+\int\omega\frac{-HR+\sqrt{1-\frac{4r_{0}}{R}+\frac{4r_{1}^{2}}{R^{2}}}}{\left(1-\frac{4r_{0}}{R}+\frac{4r_{1}^{2}}{R^{2}}
\right)^{3/2}\left(1-\frac{R^{2}}{R_{AH}^{2}}\right)}dR\right]\,,
\end{equation}
\begin{equation}\label{eq60}
\mathbf{S}_{in}\left(t,R\right)=\left[1+\sum_{n}\frac{4(1+\alpha)^{2}}{R_{AH}^{2}}\eta_{_{n}}\hbar^{n}\right]\left[-\int\omega dt+\int\omega\frac{-HR-\sqrt{1-\frac{4r_{0}}{R}+\frac{4r_{1}^{2}}{R^{2}}}}{\left(1-\frac{4r_{0}}{R}+\frac{4r_{1}^{2}}{R^{2}}
\right)^{3/2}\left(1-\frac{R^{2}}{R_{AH}^{2}}\right)}dR\right]\,.
\end{equation}

In the dynamical spacetime of the cosmological black hole solution in the STVG theory, like the FLRW universe in GR, the Hawking-like radiation is detected by the interior Kodama observer using the Kodama vector \eqref{eq24}. As previously mentioned, this is because the energy of the particle is defined by the Kodama vector, which is time-like, null, and space-like in outside, on, and inside the apparent horizon, respectively. So, there is a discrepancy between the Kodama vector of the interior and exterior regions of the apparent horizon. This discrepancy causes the temporal part of the action. Consequently, in Eqs. \eqref{eq59} and \eqref{eq60} the integral of the temporal part has also an imaginary part in the action of the tunneling particle. Such a situation can be seen in the Schwarzschild black hole \cite{Akhmedov2008} in GR. Thus, using the well-known relation
$P=\left|\psi\right|^{2}=\left|\exp\left[\frac{i}{\hbar}\mathbf{S}\left(t,R\right)\right]\right|^{2}$ in quantum mechanics, one can find the outgoing and ingoing probabilities as follows
\begin{equation}\label{eq61}
P_{out}=\exp\left[-\frac{2}{\hbar}\left(1+\sum_{n}\frac{4(1+\alpha)^{2}}{R_{AH}^{2}}\eta_{_{n}}\hbar^{n}\right)\left(-\Ima
\int\omega dt+\Ima\int\omega\frac{-HR+\sqrt{1-\frac{4r_{0}}{R}+\frac{4r_{1}^{2}}{R^{2}}}}{\left(1-\frac{4r_{0}}{R}+\frac{4r_{1}^{2}}
{R^{2}}\right)^{3/2}\left(1-\frac{R^{2}}{R_{AH}^{2}}\right)}dR\right)\right]\,,
\end{equation}
\begin{equation}\label{eq62}
P_{in}=\exp\left[-\frac{2}{\hbar}\left(1+\sum_{n}\frac{4(1+\alpha)^{2}}{R_{AH}^{2}}\eta_{_{n}}\hbar^{n}\right)\left(-\Ima
\int\omega dt+\Ima\int\omega\frac{-HR-\sqrt{1-\frac{4r_{0}}{R}+\frac{4r_{1}^{2}}{R^{2}}}}{\left(1-\frac{4r_{0}}{R}+\frac{4r_{1}^{2}}
{R^{2}}\right)^{3/2}\left(1-\frac{R^{2}}{R_{AH}^{2}}\right)}dR\right)\right]\,.
\end{equation}
Dividing $P_{out}$ by $P_{in}$ results in the tunneling rate, $\Gamma$ as follows
\begin{equation}\label{eq63}
\Gamma=\frac{P_{in}}{P_{out}} =\exp\Bigg[\frac{4}{\hbar}\omega\left(1+\sum_{n}\frac{4(1+\alpha)^{2}}{R_{AH}^{2}}\eta_{_{n}}\hbar^{n}
\right)\Ima\int\frac{1}{\left(1-\frac{4r_{0}}{R}+\frac{4r_{1}^{2}}{R^{2}}\right)\left(1-\frac{R^{2}}{R_{AH}^{2}}\right)}dR
\Bigg]\,.
\end{equation}
By computing the contour integral in Eq. \eqref{eq63} through residue theorem (for more details, see Appendix \ref{app2}), one can find the final form for the tunneling rate as follows
\begin{equation}\label{eq64}
\Gamma=\frac{P_{in}}{P_{out}}=\exp\Bigg[-\frac{2}{\hbar}\omega\left(1+\sum_{n}\frac{4(1+\alpha)^{2}}{R_{AH}^{2}}\eta_{_{n}}\hbar^{n}
\right)\frac{\pi R_{AH}}{1-\frac{4r_{0}}{R_{AH}}+\frac{4r_{1}^{2}}{R_{AH}^{2}}}\Bigg]\,.
\end{equation}
Then, the principle of ``detailed balance'' \cite{Srinivasan1999,Shankaranarayanan2002} gives us the following relation
\begin{equation}\label{eq65}
\Gamma=\frac{P_{in}}{P_{out}}=\exp\left[-\frac{\omega}{T_{_{BSA}}}\right]\,,
\end{equation}
where `BSA' stands for beyond the semiclassical approximation. Finally, we can find the Hawking-like temperature corresponding with the apparent horizons in the cosmological black hole solution within the STVG framework beyond the semiclassical approximation as follows
\begin{equation}\label{eq66}
T_{_{BSA}}=T\left(1+\sum_{n}\frac{4(1+\alpha)^{2}}{R_{AH}^{2}}\eta_{_{n}}\hbar^{n}\right)^{-1}\,,\quad T=\frac{\hbar}{2\pi R_{AH}}\left(1-\frac{2M(1+\alpha)}{R_{AH}}+\frac{M^{2}\alpha(1+\alpha)}{R_{AH}^{2}}\right)\,,
\end{equation}
in which $T$ is the semiclassical Hawking-like temperature, which we found in the previous section in Eqs. \eqref{eq35} and \eqref{eq46}. Also, the other terms are the quantum corrections arising from the higher-order quantum effects due to the presence of $\hbar$. Accordingly, we have found all the quantum corrections to the semiclassical Hawking-like temperature derived in the previous section, associated with the thermal Hawking-like radiation of apparent horizons of the cosmological black hole in the STVG theory. Again, all deduced results in the limit of $\alpha\rightarrow 0$ tend to the corresponding ones in the McVittie solution in GR and hence it is consistent with correspondence principle. Additionally, in the limit of $\hbar\rightarrow 0$ the semiclassical Hawking-like temperature will recover, i.e., $\lim_{\hbar\rightarrow 0}T_{_{BSA}}=T$. So, it should be noted that the quantum corrections are due to the presence of $\hbar$ and also, in the expansion of $\mathbf{S}$ there is no need to consider whether $\mathbf{S}$ and $\mathbf{S}_{n}$ are real or imaginary.

\section{Summary and Conclusions}\label{SaC}

In this paper, we took into account the first cosmological black hole solution within the weak field regime of the STVG theory, which lives in the dynamical FLRW background \cite{Perez2019}. While cosmological black holes in other modified gravity theories in static limit reduce to naked singularities, we noticed that the solution in STVG theory in such a limit reduces to the Schwarzschild-MOG black hole. Hence, one can conclude that STVG theory is more suitable to explain black holes both on local and global length scales, which is another capability of this classical theory to describe various manifestations of gravity. We saw that the cosmological black hole in the STVG theory has three apparent horizons: a cosmological apparent horizon, $R_{+}$ which approaches the cosmological apparent horizon of the FLRW universe, a cosmological event horizon, $R_{-}$ which tends to the singularity, and an apparent horizon, $R_{*}$ within the cosmological singularity disconnected from exterior geometry. By plotting these apparent horizons in terms of cosmic time for different values of the dimensionless STVG parameter, $\alpha$ we showed that for larger values of $\alpha$ the cosmological apparent and event horizons emerge together at some later cosmic time.

In the present paper, we proved in details that the apparent horizons of the cosmological black hole solution in the STVG theory has indeed an ingoing Hawking-like radiation. By doing so, our main goals were to see how the STVG theory and its dimensionless parameter, $\alpha$ affect the Hawking-like radiation, and also, how the corresponding temperature will become a function of the parameter as a discriminant tool between STVG and GR. In this way, we followed the Hamilton-Jacobi tunneling method for massive particles without back-reaction effects, and the Parikh-Wilczek tunneling approach for massless particles in the presence of back-reaction effects, both based on semiclassical approximation. We proved that the back-reaction effects in the Parikh-Wilczek method have led to addressing the information paradox by a non-zero correlation function, strongly dependent on the parameter $\alpha$ of the STVG theory. On the other hand, we used the Hamilton-Jacobi method beyond the semiclassical approximation for a tunneling massless scalar field to involve all higher-order quantum corrections in the single particle action, and we showed that these higher-order corrections terms are proportional to the semiclassical contribution. Also, based on the semiclassical and beyond semiclassical approximation, we showed that such an ingoing Hawking-like radiation associated with apparent horizons of the cosmological black hole solution is measured by an interior Kodama observer with a Kodama vector in the dynamical setup. Therefore, the Hawking effect is not only for event horizons of stationary black holes but also for apparent horizons of non-stationary black holes, even in STVG theory. Moreover, we proved that the outgoing tunneling rate is never available in the setup. We found out the emission rate of the ingoing Hawking-like radiation and the corresponding Hawking-like temperature of the apparent horizons $R_{-}$ and $R_{+}$ of the cosmological black hole in the STVG theory in both Hamilton-Jacobi and Parikh-Wilczek methods based on the semiclassical approximation. Then, we plotted the natural logarithm's function of the emission rate, $\ln\,[\Gamma]$ in the Hamilton-Jacobi method based on the semiclassical approximation in terms of cosmic time for different values of $\alpha$. We concluded that the larger the values of $\alpha$, the later this function becomes available. This is also true in the case of Parikh-Wilczek method with back-reaction effects. Also, by plotting the corresponding Hawking-like temperature in the Hamilton-Jacobi method based on the semiclassical approximation in terms of cosmic time for different values of $\alpha$, we showed that the larger the values of $\alpha$, the later the Hawking-like temperature becomes available with some smaller values \cite{Cai2021,Mureika2016}. This is also true in the cases of Parikh-Wilczek method based on the semiclassical approximation with back-reaction effects and the Hamilton-Jacobi approach beyond the semiclassical approximation for which the temperature plotted in terms of MSH mass for different values of $\alpha$. Additionally, all the deduced results for the cosmological black hole in the STVG theory completely satisfied the correspondence principle so that for $\alpha=0$ they became the Hawking-like temperature of the apparent horizons of McVittie spacetime, which was not reported in the literature and also, by setting $\alpha=0\,,\, M=0$ they became the Hawking-like temperature of the apparent horizon of the spatially flat FLRW universe \cite{Cai2009}. This work can be a step towards a better understanding of the cosmological black hole in the STVG setup and also, the behavior of the Hawking-like radiation of apparent horizons in the theory. In future work, we aim to study the thermodynamic of the cosmological black hole in the STVG framework to check if there is indeed a deep connection between laws of thermodynamic, especially unified first law of thermodynamics \cite{Faraoni2015} and STVG field equations.

Finally, the resulted Hawking temperature in this non-stationary spacetime does not obey equilibrium distribution. This is because the resulted temperature is not a constant physical quantity; instead, it is a function of time coordinate. Therefore, this Hawking-like temperature alters over time, and the interior Kodama observer measures different values of the time-dependent temperature at every moment of time. Consequently, the present system is in a non-equilibrium situation, as mentioned before. Whereas, the exact definition of temperature and thermodynamic laws in a non-equilibrium situation is not found, completely. In fact, for the systems in non-equilibrium distribution, any thermometer with sensitivity to distinct degrees of freedom will measure different values of temperature. This makes it difficult to define temperature in these non-equilibrium systems, uniquely \cite{Casas2003}. Considering some defined degrees of freedom, however, the temperature can be described with respect to confined correctness of the zeroth law of thermodynamics (associated with these defined degrees of freedom) in non-equilibrium distribution. This leads to ``local'' (effective) definition of temperature and thermodynamic quantities in such systems \cite{Casas2003}. This local definition of temperature is based on the near-horizon (local) nature of the Hawking (-like) radiation \cite{Bhattacharya2016}. The same concept is also valid in our study for the resulted time-dependent temperature of the cosmological black hole. Such effective temperature is also determined by S. Weinberg \cite{Casas2003} for a non-equilibrium system of photons to the form of $$\frac{\Omega}{\Lambda}=e^{-\frac{E}{T_{eff}}}$$ where $\Lambda$ is the absorption rate coefficient and $\Omega$ is the stimulated emission coefficient. On the other hand, the transmission coefficient $\Gamma$ in our work has become similar to Boltzmann factor, as same as the outcome of Refs. \cite{Bhattacharya2016,Dalui2021}. To be precise, however, the above relation argued by S. Weinberg \cite{Casas2003} is implemented to find the effective temperature in non-equilibrium situation. In other words, for equilibrium situations, the exact Boltzmann factor is used to determine the global temperature, while in time-dependent cases, only an approximation of particle action and transmission coefficient, called near-horizon approximation, has become similar to Boltzmann factor (for more details, see, e.g. \cite{Bhattacharya2016,Dalui2021,Casas2003,Vanzo2011}).

\begin{acknowledgments}

The authors would like to thank John W. Moffat for fruitful comments and discussions. 

\end{acknowledgments}

\appendix

\section{Derivation of the radial null geodesic in Equation \eqref{eq36}}\label{app1}

Eq. \eqref{eq36} describes the radial null geodesic of the cosmological black hole solution in STVG theory in PPG coordinates. To derive this equation, we first recall the line element of the cosmological black hole solution in STVG theory characterized by the PPG coordinates system in \eqref{eq23}
\begin{equation}\label{ap1}
ds^{2}=-\left(1-\frac{4r_{0}}{R}+\frac{4r_{1}^{2}}{R^{2}}-H^{2}R^{2}\right)dt^{2}-\frac{2RHdR dt}{\sqrt{1-\frac{4r_{0}}{R}+\frac{4r_{1}^{2}}{R^{2}}}}
+\frac{dR^{2}}{1-\frac{4r_{0}}{R}+\frac{4r_{1}^{2}}{R^{2}}}+R^{2}d\Omega^{2}\,.
\end{equation}
For radial null geodesic, we have $ds^{2}=d\Omega^{2}=0$ which results in the following relation
\begin{equation}\label{ap2}
0=-\left(1-\frac{4r_{0}}{R}+\frac{4r_{1}^{2}}{R^{2}}-H^{2}R^{2}\right)dt^{2}-\frac{2RHdR dt}{\sqrt{1-\frac{4r_{0}}{R}+\frac{4r_{1}^{2}}{R^{2}}}}
+\frac{dR^{2}}{1-\frac{4r_{0}}{R}+\frac{4r_{1}^{2}}{R^{2}}}\,.
\end{equation}
Dividing both sides of Eq. \eqref{ap2} by $dt^{2}$ one can write
\begin{equation}\label{ap3}
0=-\left(1-\frac{4r_{0}}{R}+\frac{4r_{1}^{2}}{R^{2}}-H^{2}R^{2}\right)-\frac{2RH\dot{R}}{\sqrt{1-\frac{4r_{0}}{R}
+\frac{4r_{1}^{2}}{R^{2}}}}+\frac{\dot{R}^{2}}{1-\frac{4r_{0}}{R}+\frac{4r_{1}^{2}}{R^{2}}}\,,
\end{equation}
where dot stands for time derivative. Now, Eq. \eqref{ap3} is a quadratic equation for $R$, which its solutions are as follows
\begin{equation}\label{ap4}
\frac{dR}{dt}=\dot{R}=\pm\sqrt{1-\frac{4r_{0}}{R}+\frac{4r_{1}^{2}}{R^{2}}}\left(\sqrt{1-\frac{4r_{0}}{R}+\frac{4r_{1}^{2}}
{R^{2}}}\pm HR\right)\,.
\end{equation}
So, Eq. \eqref{ap4} is the radial null geodesic in Eq. \eqref{eq36}

\section{The poles of contour integrals in calculating tunneling and emission rates}\label{app2}

The integrals in Eqs. \eqref{eq32}, \eqref{eq33}, and \eqref{eq63} have four poles as follows
\begin{equation}\label{ap5}
\begin{split}
& R_{1}=+R_{AH}\,,\quad\quad\quad\quad\quad\quad\,\,\,\,\,\, R_{2}=-R_{AH}\,, \\
& R_{3}=2\left(r_{0}+\sqrt{r_{0}^{2}-r_{1}^{2}}\right)\,,\qquad R_{4}=2\left(r_{0}-\sqrt{r_{0}^{2}-r_{1}^{2}}\right)\,.
\end{split}
\end{equation}
As we mentioned in Section \ref{STVG}, the spacetime events located in the casual future of the cosmological singularity are interested in this study. Thus, we focus only on the cosmological apparent horizon $R_{+}$, and the cosmological event horizon $R_{-}$. As seen from Figs.\ref{Fig1} and \ref{Fig2}, $R_{+}$ and $R_{-}$ are always positive over cosmic time. Therefore, the pole located at $R_{2}=-R_{AH}$ is negative and so, obviously is not physical, since $R_{AH}$ associated with $R_{+}$ and $R_{-}$ is positive. So, the pole $R_{2}$ do not affect the results of Eqs. \eqref{eq32}, \eqref{eq33}, and \eqref{eq63}. On the other hand, from Eq. \eqref{eq15}, one can find that the last two poles can be rewritten as
\begin{equation}\label{ap6}
\begin{split}
& R_{3}=2\left(r_{0}+\sqrt{r_{0}^{2}-r_{1}^{2}}\right)=M\left(1+\alpha+\sqrt{1+\alpha}\right)\,,\\
& R_{4}=2\left(r_{0}-\sqrt{r_{0}^{2}-r_{1}^{2}}\right)=M\left(1+\alpha-\sqrt{1+\alpha}\right)\,.
\end{split}
\end{equation}
\begin{figure}[htb]
\centering
\includegraphics[width=0.7\textwidth]{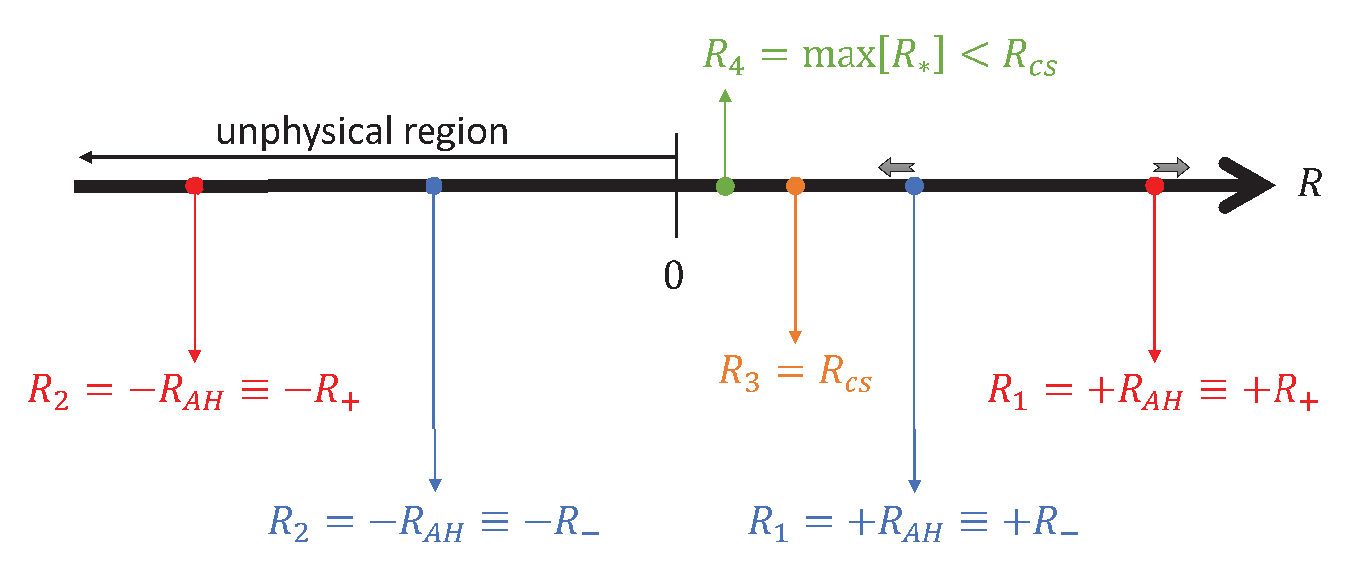}
\caption{\label{Fig6}\small{\emph{A graphic to show the location of these four poles over areal radius, $R$. The region $R<0$ is the unphysical region at which the pole $R_{2}=-R_{AH}$ is located, and $R_{AH}$ can be either $R_{-}$ or $R_{+}$. At the physical region $R>0$, three other poles, i.e., $R_{1}$, $R_{3}$, and $R_{4}$ are located. $R_{3}$ coincides with the cosmological singularity of the black hole and $R_{4}$ equals to the maximum value of $R_{*}$. The pole $R_{1}=+R_{AH}$ is the only contributed pole in calculating $\Gamma$. The evolution of the $R_{-}$ and $R_{+}$ is shown by two arrows indicating the direction of their approaching, so that in the limit of $t\rightarrow\infty$, the $R_{-}$ tends to the cosmological singularity and $R_{+}$ tends to the cosmological apparent horizon in the FLRW spacetime.}}}
\end{figure}
From Eq. \eqref{eq16} we see that $R_{3}=R_{cs}$ in which $R_{cs}$ is the cosmological singularity of the cosmological black hole in STVG setup. Therefore, one can clearly see that $R_{4}<R_{cs}$, i.e., $R_{4}$ is somewhere within the cosmological singularity. One can show that $R_{4}=R_{*}\left|\right._{t\rightarrow\infty}$ which equals to the maximum value of $R_{*}$ i.e., $R_{4}=R_{*}\left|\right._{t\rightarrow\infty}=\max\left[R_{*}\right]$. Fig.\ref{Fig6} is a graphic diagram to illustrate the location of these four poles over areal radius, $R$. So, the poles $R_{3}$ and $R_{4}$ are not within the contour of the integration in Eqs. \eqref{eq32}, \eqref{eq33}, and \eqref{eq63} and do not contribute to the residue theorem. Consequently, it is clear that only the pole located at $R_{1}=+R_{AH}$ contributes to the final results of Eqs. \eqref{eq32}, \eqref{eq33}, and \eqref{eq63} as we took it into account in deriving their results. Thus, in the mathematical point of view, there are four poles in Eqs. \eqref{eq32}, \eqref{eq33}, and \eqref{eq63}. In the physical sight, however, in our setup only the pole located at $R=+R_{AH}$ affect the tunneling rate $\Gamma$. Also, we should note that the poles $R_{3}$ and $R_{4}$ are exist in Eq. \eqref{eq44}, which again do not affect the final result of the equation, based on the above explanation. The contributed pole in Eq. \eqref{eq44} is located at $R=R_{AH}+\omega$ due to the presence of back-reaction effects.

\end{document}